\documentclass[twocolumn]{aastex62}

\received{\today}
\revised{}
\accepted{}
\submitjournal{ApJ}

\shorttitle{Observation of three-dimensional reconnection geometries in solar eruptions}
\shortauthors{Dud\'{i}k et al.}

\begin{document}

\title{Observation of all pre- and post-reconnection structures \\ involved in
three-dimensional reconnection geometries in solar eruptions}

\correspondingauthor{Jaroslav Dud\'{i}k}
\email{dudik@asu.cas.cz}

\author[0000-0003-1308-7427]{Jaroslav Dud\'{i}k}
\affil{Astronomical Institute of the Czech Academy of Sciences, Fri\v{c}ova 298, 251 65 Ond\v{r}ejov, Czech Republic}

\author[0000-0002-9690-8456]{Juraj L\"{o}rin\v{c}\'{i}k}
\affil{Astronomical Institute of the Czech Academy of Sciences, Fri\v{c}ova 298, 251 65 Ond\v{r}ejov, Czech Republic}

\author[0000-0001-5810-1566]{Guillaume Aulanier}
\affiliation{LESIA, Observatoire de Paris, Psl Research University, CNRS, Sorbonne Universités, UPMC Univ. Paris 06, Univ. Paris Diderot, Sorbonne Paris Cité, 5 place Jules Janssen, F-92195 Meudon, France}

\author[0000-0002-7565-5437]{Alena Zemanov\'{a}}
\affil{Astronomical Institute of the Czech Academy of Sciences, Fri\v{c}ova 298, 251 65 Ond\v{r}ejov, Czech Republic}

\author[0000-0003-3364-9183]{Brigitte Schmieder}
\affiliation{LESIA, Observatoire de Paris, Psl Research University, CNRS, Sorbonne Universités, UPMC Univ. Paris 06, Univ. Paris Diderot, Sorbonne Paris Cité, 5 place Jules Janssen, F-92195 Meudon, France}

\begin{abstract}

We report on observations of the two newly-identified reconnection geometries involving erupting flux ropes. In 3D, a flux rope can reconnect either with a surrounding coronal arcade (recently named ``ar--rf'' reconnection) or with itself (``rr--rf'' reconnection), and both kinds of reconnection create a new flux rope field line and a flare loop. For the first time, we identify all four constituents of both reconnections in a solar eruptive event, the filament eruption of 2011 June 07 observed by \textit{SDO}/AIA. The ar--rf reconnection manifests itself as shift of one leg of the filament by more than 25$\arcsec$ northward. At its previous location, a flare arcade is formed, while the new location of the filament leg previously corresponded to a footpoint of a coronal loop in 171\,\AA. In addition, the evolution of the flare ribbon hooks is also consistent with the occurrence of ar--rf reconnection as predicted by MHD simulations. Specifically, the growing hook sweeps footpoints of preeruptive coronal arcades, and these locations become inside the hook. Furthermore, the rr--rf reconnection occurs during the peak phase above the flare arcade, in an apparently X-type geometry involving a pair of converging bright filament strands in the erupting filament. A new flare loop forms near the leg of one of the strands, while a bright blob, representing a remnant of the same strand, is seen ascending into the erupting filament. All together, these observations vindicate recent predictions of the 3D standard solar flare model.

\end{abstract}


%
\section{Introduction}
\label{Sect:1}

Solar flares and eruptions are the brightest and the most violent manifestations of solar magnetic activity \citep{Fletcher11,Schmieder15}. The well-known 2D CSKHP standard solar flare model \citep{Carmichael64,Sturrock66,Hirayama74,Kopp76} was recently extended into three dimensions (3D) by \citet{Aulanier12,Aulanier13} and \citet{Janvier13,Janvier15}. The 3D model successfully underwent a series of observational tests and explained many inherently 3D flare phenomena, such as the shape of flare ribbons and their hooks, relationship of these J-shaped ribbons to electric currents \citep{Janvier14} and apparent slipping motion of flare loops \citep{Dudik14a,Dudik16,Li14,Li15,Li16}, as well as the existence of the vortices at the flanks of the erupting magnetic flux rope \citep{Zuccarello17,Dudik17a}, and eruption-induced changes in the surface magnetic field \citep{Barczynski19}.

Recently, \citet{Aulanier19} found that, in addition to the standard reconnection between coronal arcades \textit{a}, which builds the flux rope \textit{r} and creates flare loops \textit{f} as described in the CSHKP model, denoted as the {aa--rf} reconnection, new reconnection geometries can also exist in three dimensions. The erupting flux rope \textit{r} also undergoes reconnections with the surrounding coronal arcades \textit{a}, leading to creation of new rope field lines \textit{r} and flare loops \textit{f}. This purely 3D \textit{ar--rf} reconnection geometry erodes the flux rope on the inner side, where the straight part of the flare ribbon curves into its hooked section \citep[see Figure 5 in][]{Aulanier19}. This erosion of the rope on the inner side happens as the straight parts of the ribbons move outward. Simultaneously, the hooks of the ribbons drift, and the coronal arcades located previously on the outer edges of the hooks become new flux rope field lines. This process leads to drifting of the flux rope footpoints during the course of the eruption, in spite of the line-tying of the flux rope footpoints \citet{Aulanier19}. Finally, the existence of the ar--rf reconnection means that flare loops rooted at the ends of the straight parts of J-shaped ribbons are \textit{not} formed by the same reconnection geometry as the remainder of flare loops within the same flare arcade, whose genesis is the standard aa--rf reconnection between coronal arcades.

In addition to that, \citet{Aulanier19} predicted the existence of the \textit{rr--rf} reconnection: When the eruption is well underway, the internal, near-vertical legs of the erupting magnetic flux rope reconnect to create a new, multi-turn flux rope field line and a flare loop. The footpoints of the field lines involved in the  rr--rf reconnection are located near the tips of the flare ribbon hooks. The rr--rf reconnection leads to increase of the poloidal flux of the erupting flux rope. 

While the standard aa--rf reconnection has been observed in several events either as coronal inflows \citep[e.g.,][]{Yokoyama01,Zhu16} or as apparently slipping flare loops \citep[e.g.,][]{Dudik14a}, the only evidence for ar--rf reconnection as presented in \citet{Aulanier19} relied only on ribbon deformation. In the works of \citet{Zemanova19} and \citet{Lorincik19}, the evidence was incomplete, as only some of the four components of the ar--rf reconnection were visible. In \citet{Lorincik19}, footpoints of an erupting filament were swept by the ribbon hook and turned to flare loops, a signature of the ar--rf reconnection. \citet{Zemanova19} observed significant drift of a ribbon hook in another event, followed by complex evolution of this hook, consistent with the ar--rf reconnection. As for the rr--rf reconnection, to our knowledge, the evidence for this reconnection geometry is wholly missing. Here, we report on the occurrence of both ar--rf and rr--rf reconnections in the filament eruption of 2011 June 07. The event is presented in Section \ref{Sect:2}. Sections \ref{Sect:3} and \ref{Sect:4} detail observations of the ar--rf and rr--rf reconnections, respectively. Conclusions are given in Section \ref{Sect:5}.

\begin{figure*}[ht]
	\centering 

	\includegraphics[width=16.00cm,viewport= 0  0 997 260,clip]{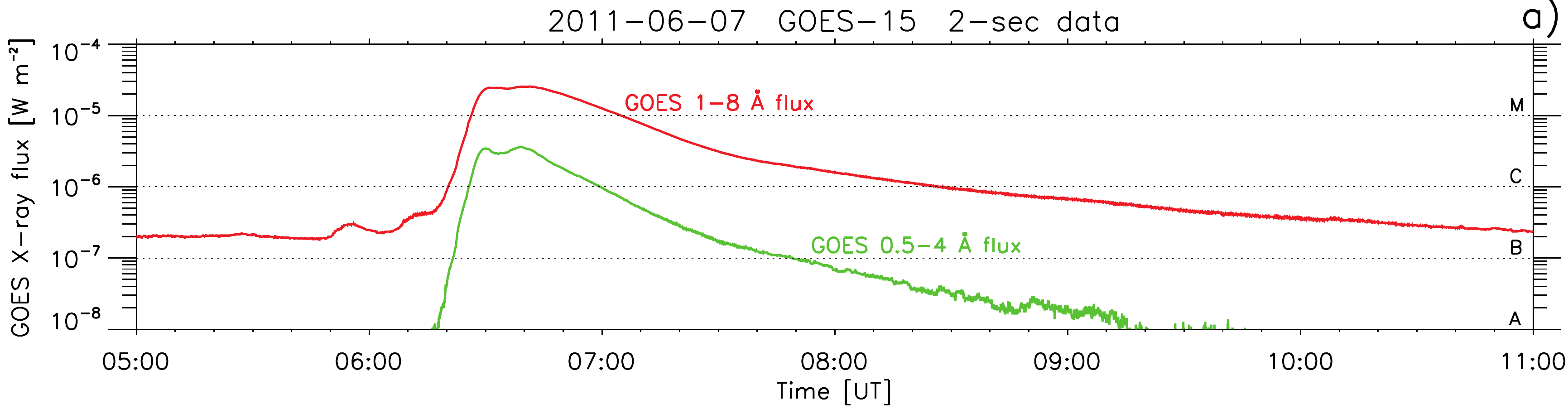}
	\includegraphics[width=8.61cm,viewport= 0 45 495 355,clip]{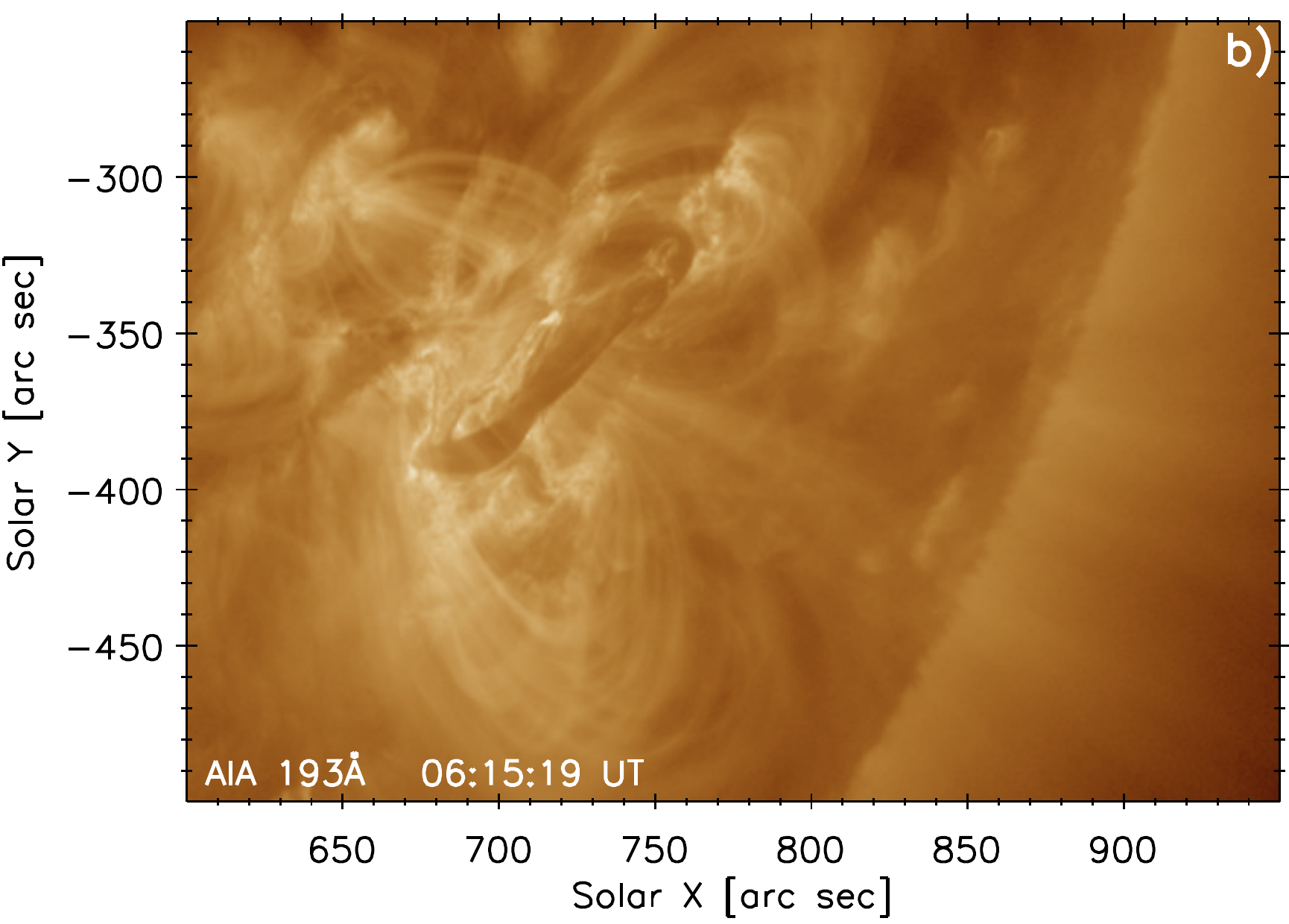}
	\includegraphics[width=7.39cm,viewport=70 45 495 355,clip]{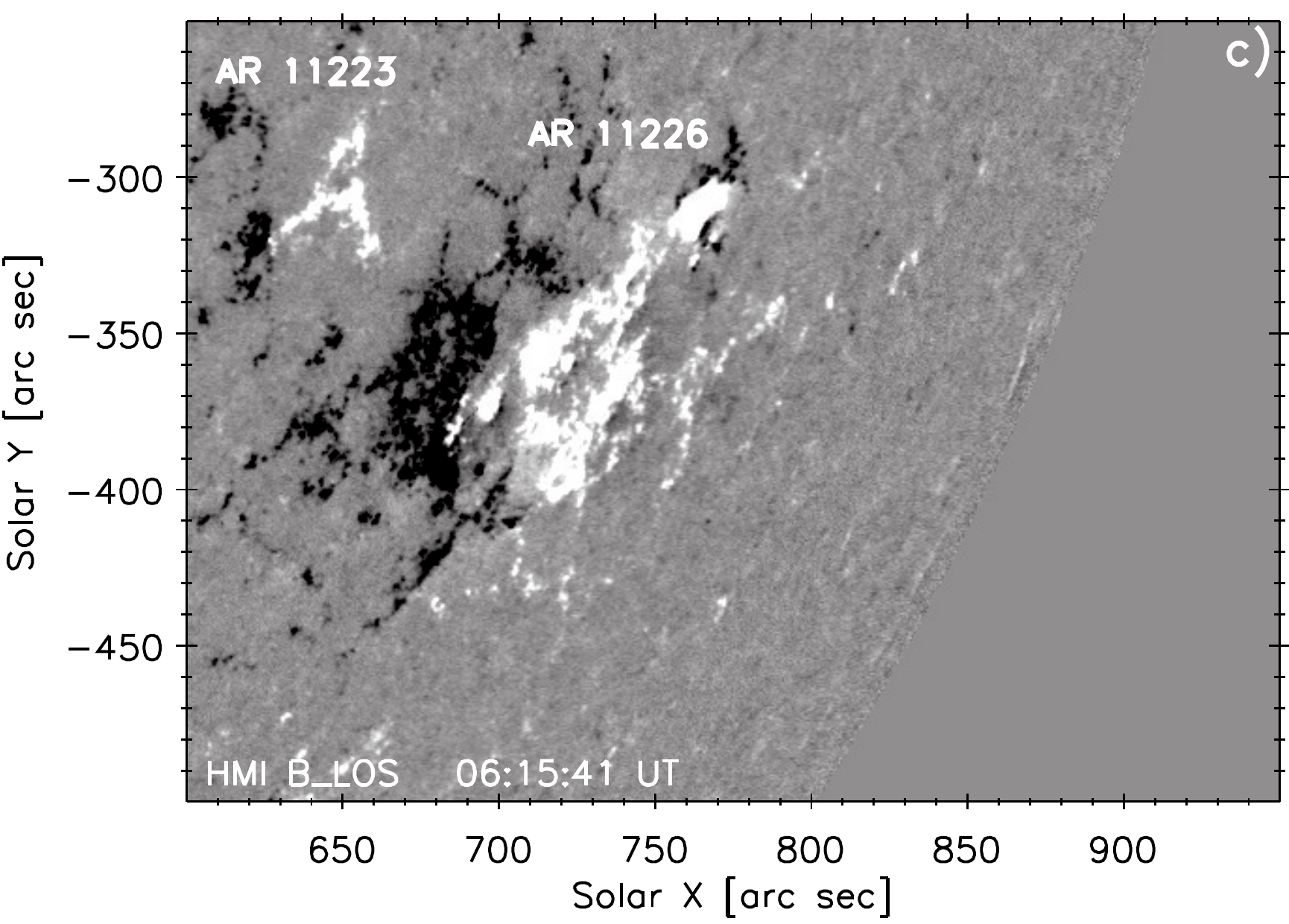}
	\includegraphics[width=8.61cm,viewport= 0 45 495 355,clip]{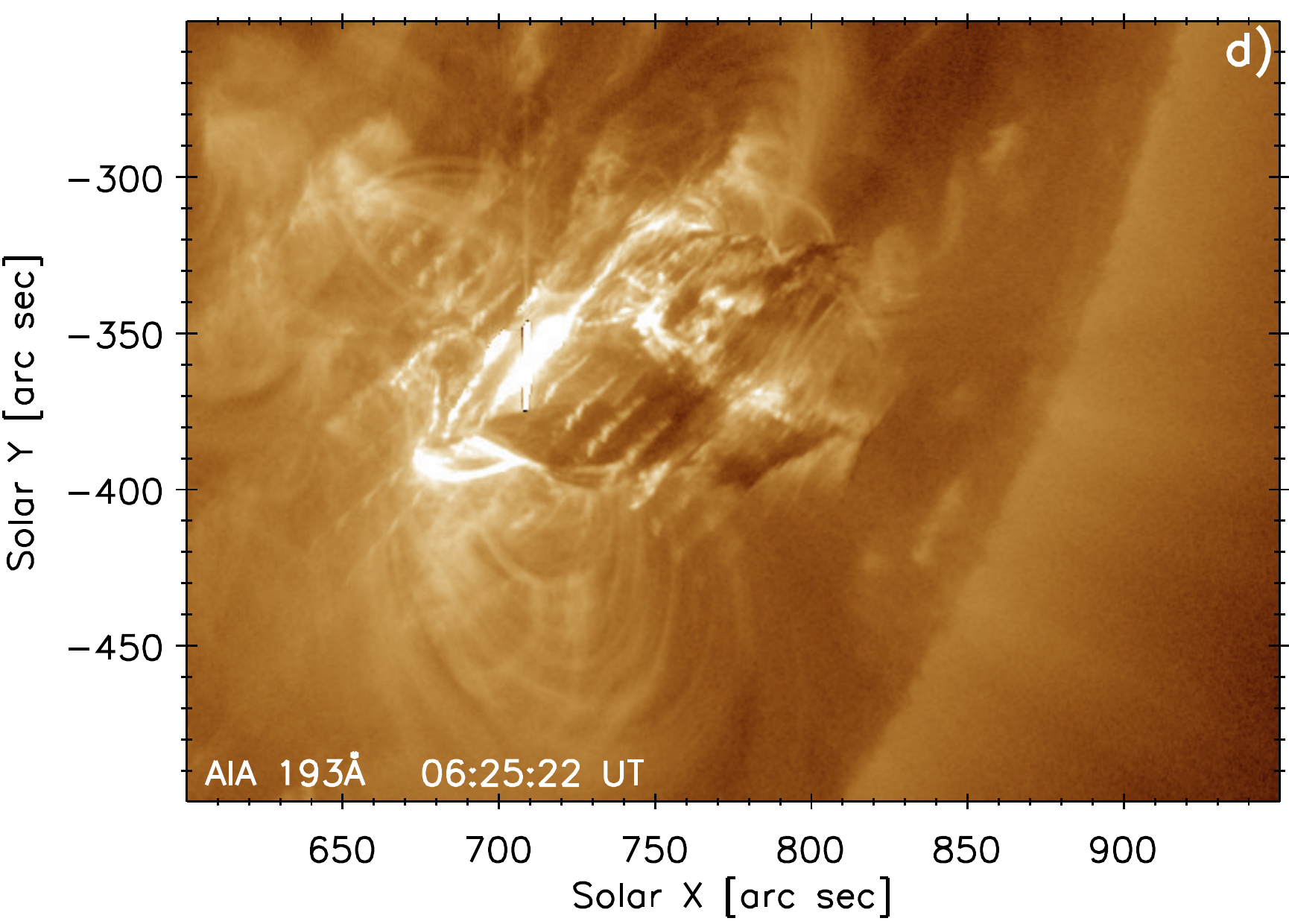}
	\includegraphics[width=7.39cm,viewport=70 45 495 355,clip]{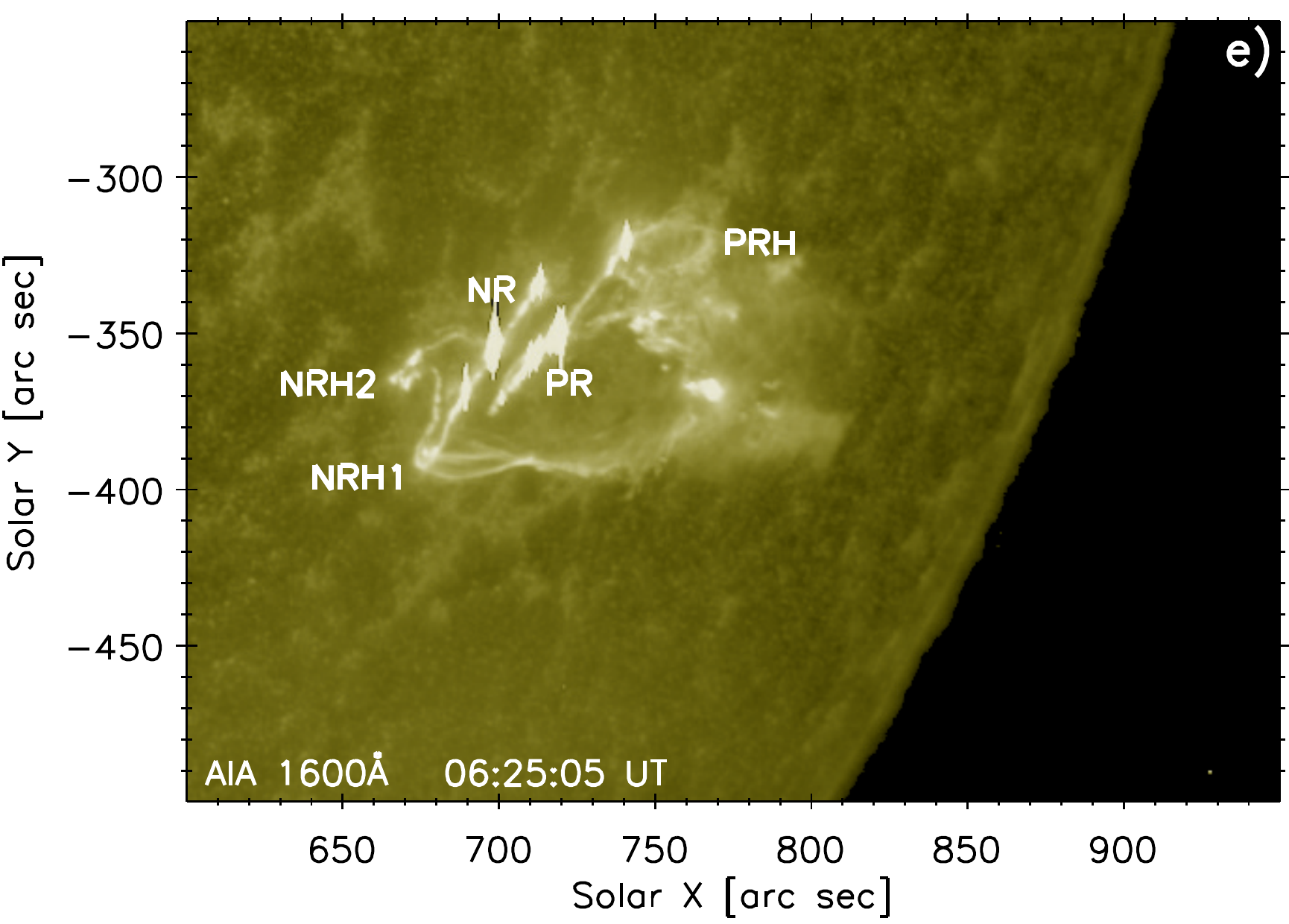}
	\includegraphics[width=8.61cm,viewport= 0  0 495 355,clip]{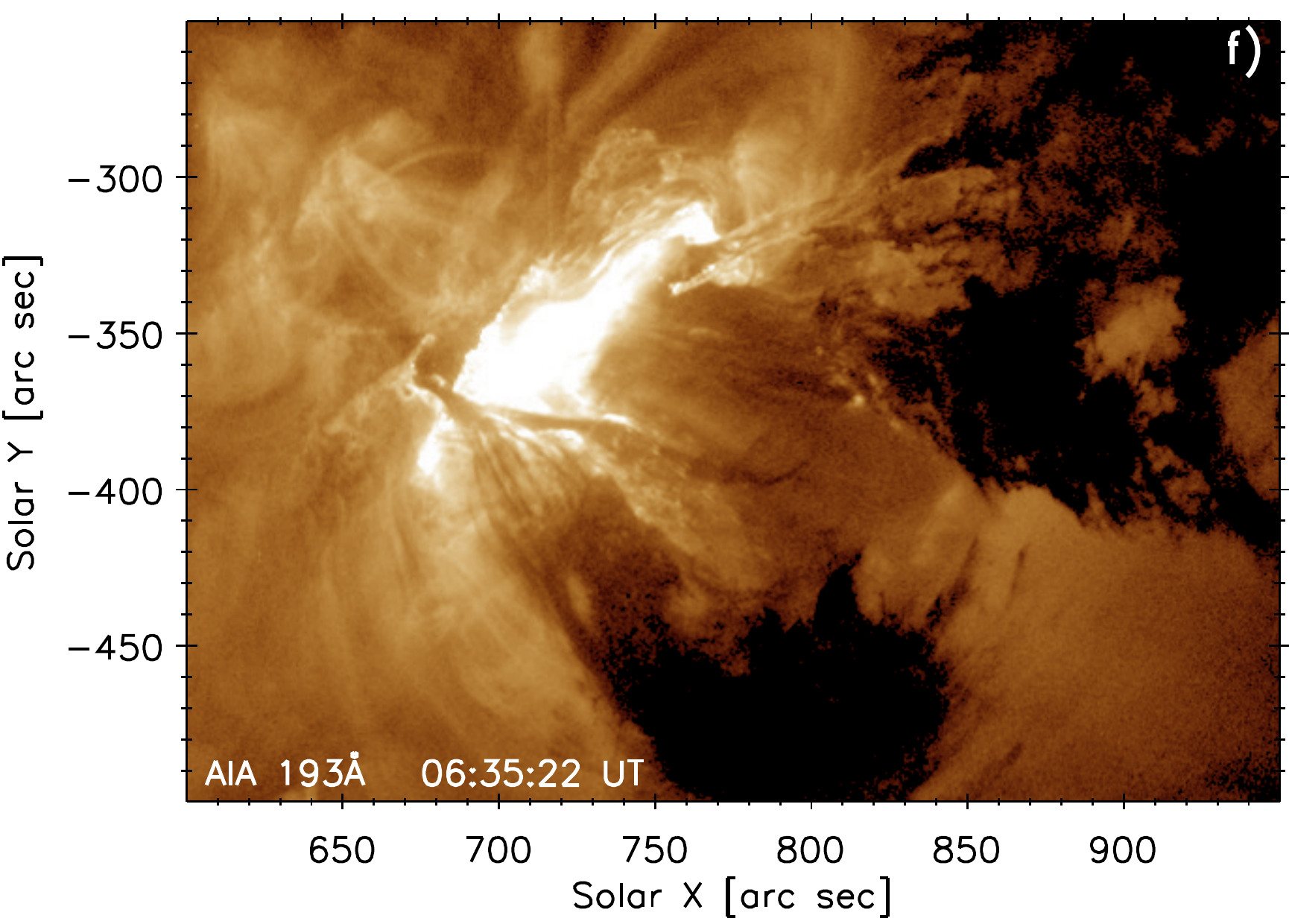}
	\includegraphics[width=7.39cm,viewport=70  0 495 355,clip]{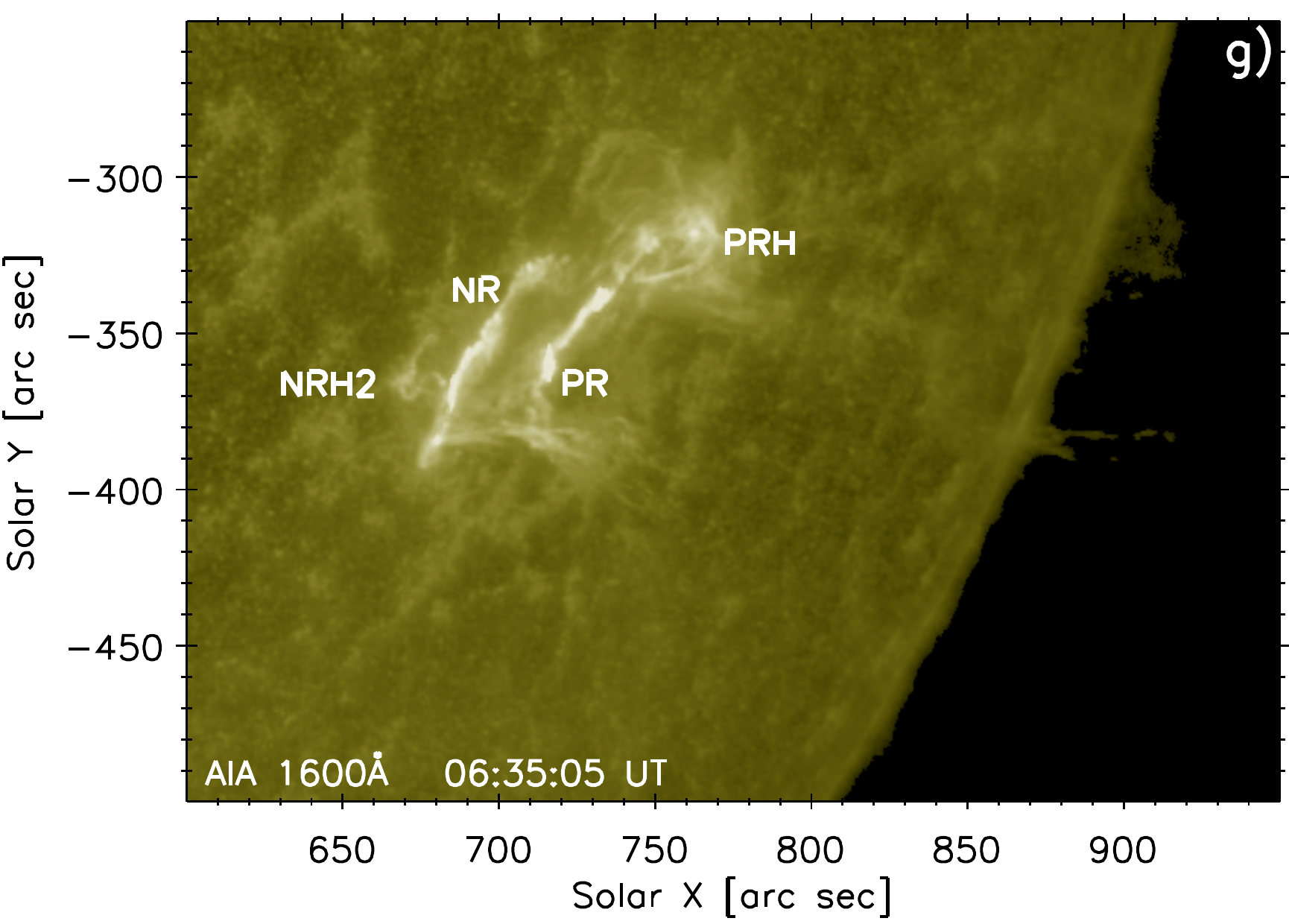}
\caption{Overview of the filament eruption of 2011 June 07 and the accompanying flare. (a): Evolution of the \textit{GOES} X-ray flux. (b, d, f): Eruption as observed by the \textit{SDO}/AIA 193\,\AA~channel. Bright flare ribbons are evident, and shown for clarity in panels (e) and (g). Panel (c) shows the corresponding line-of-sight component of the magnetic field from \textit{SDO}/HMI.}
\label{Fig:Eruption}
\end{figure*}

\begin{figure*}
	\centering
	\includegraphics[width=5.32cm,viewport= 0  0 282 278,clip]{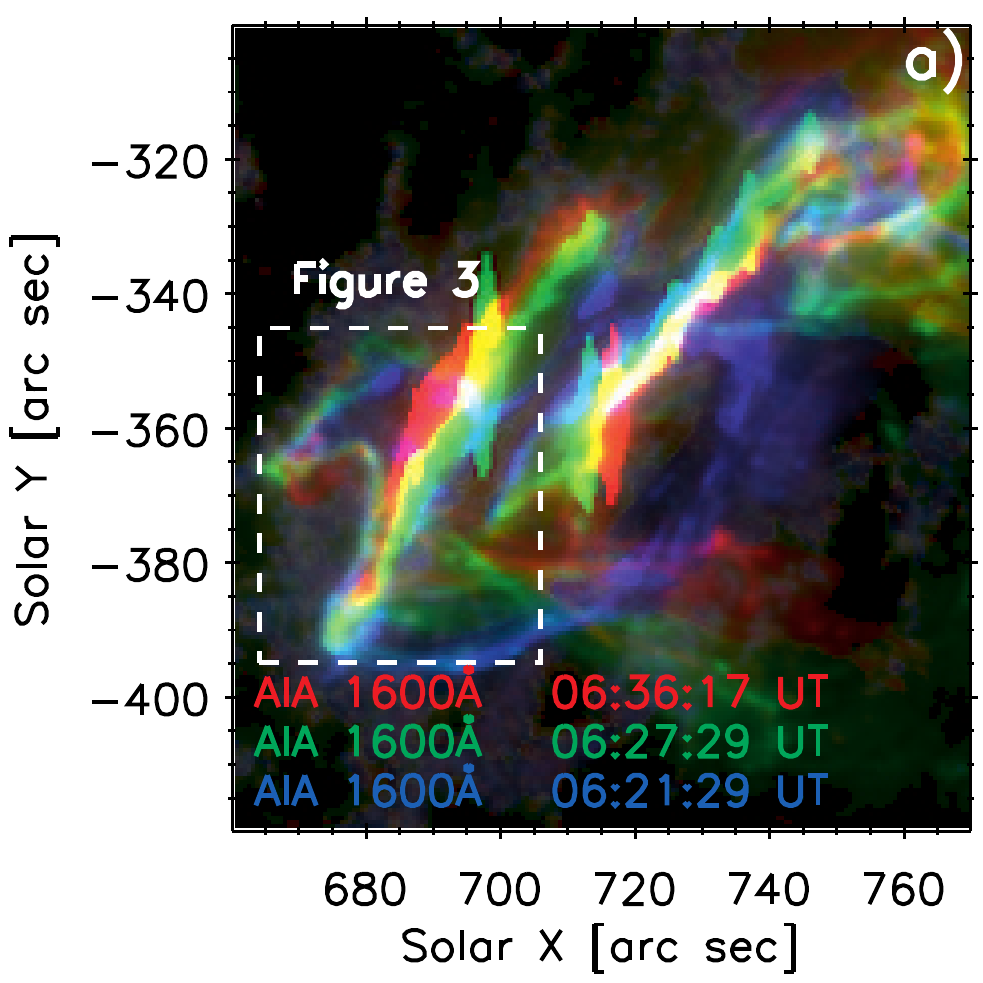}
	\includegraphics[width=4.10cm,viewport=65  0 282 278,clip]{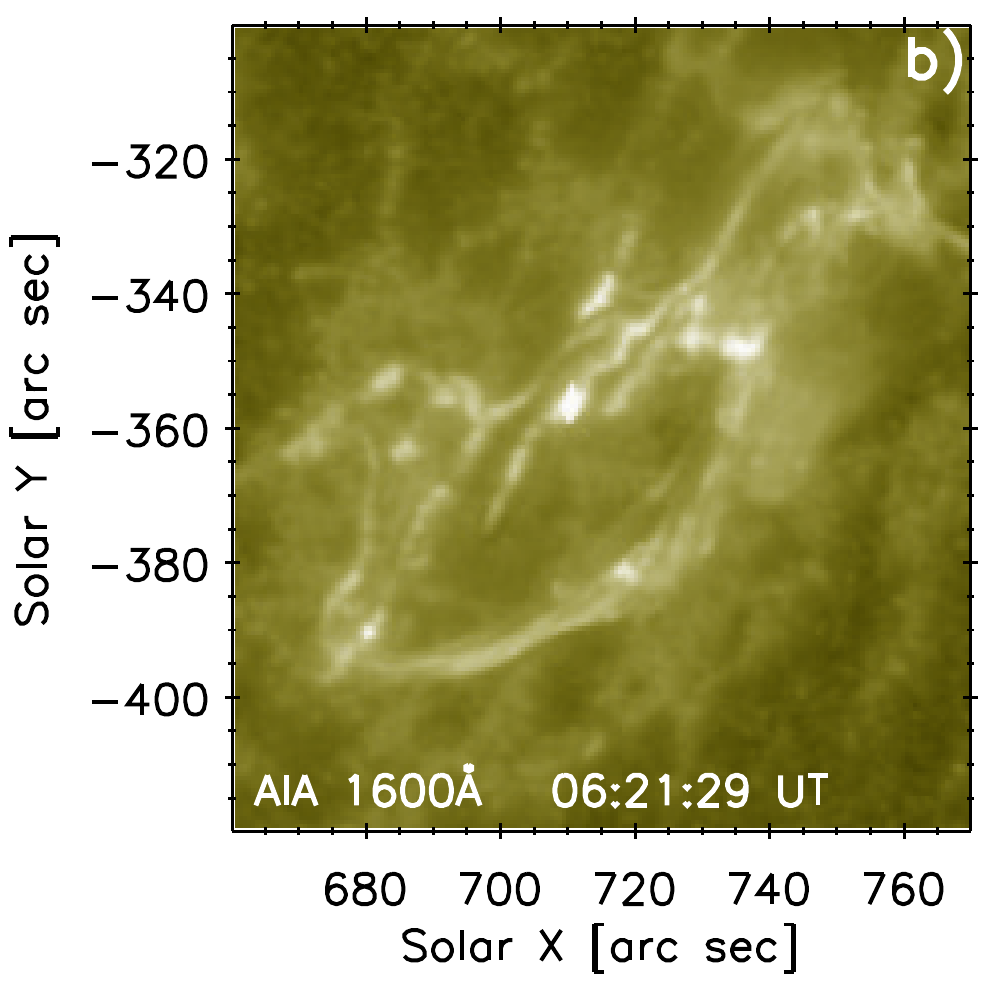}
	\includegraphics[width=4.10cm,viewport=65  0 282 278,clip]{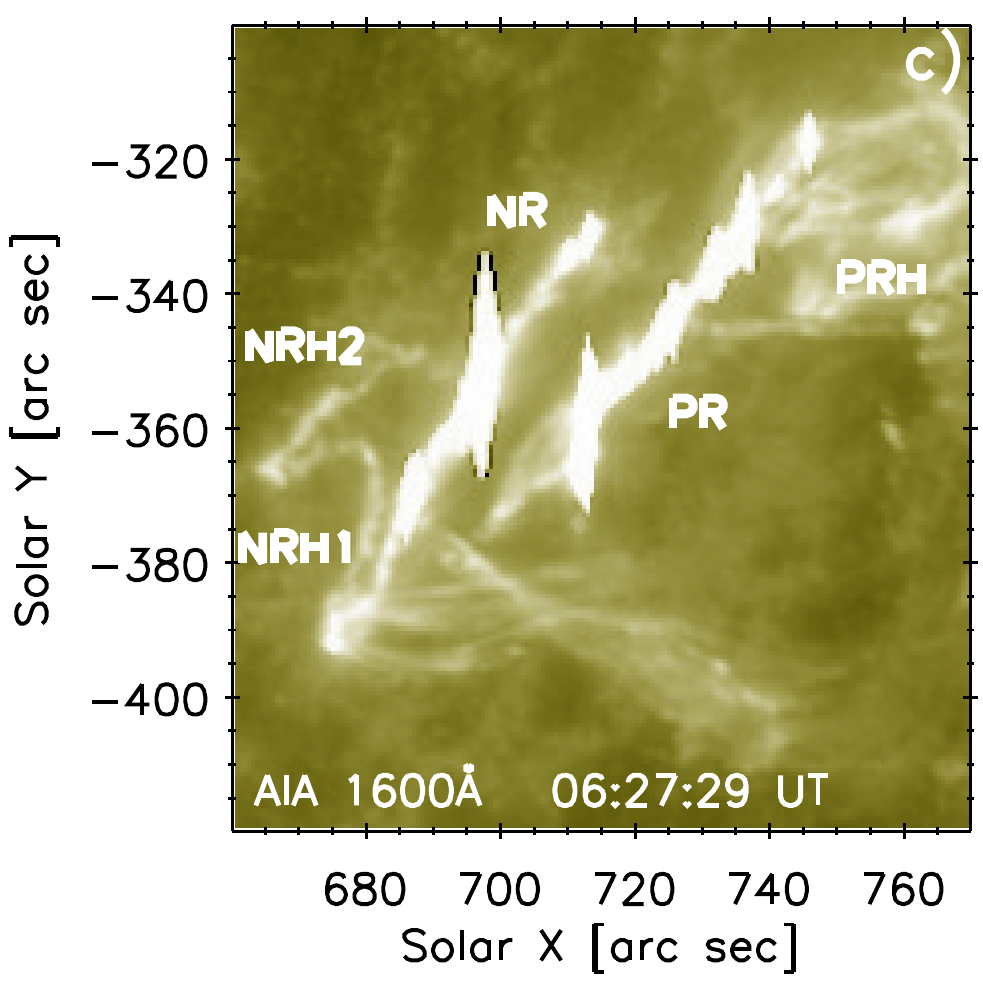}
	\includegraphics[width=4.10cm,viewport=65  0 282 278,clip]{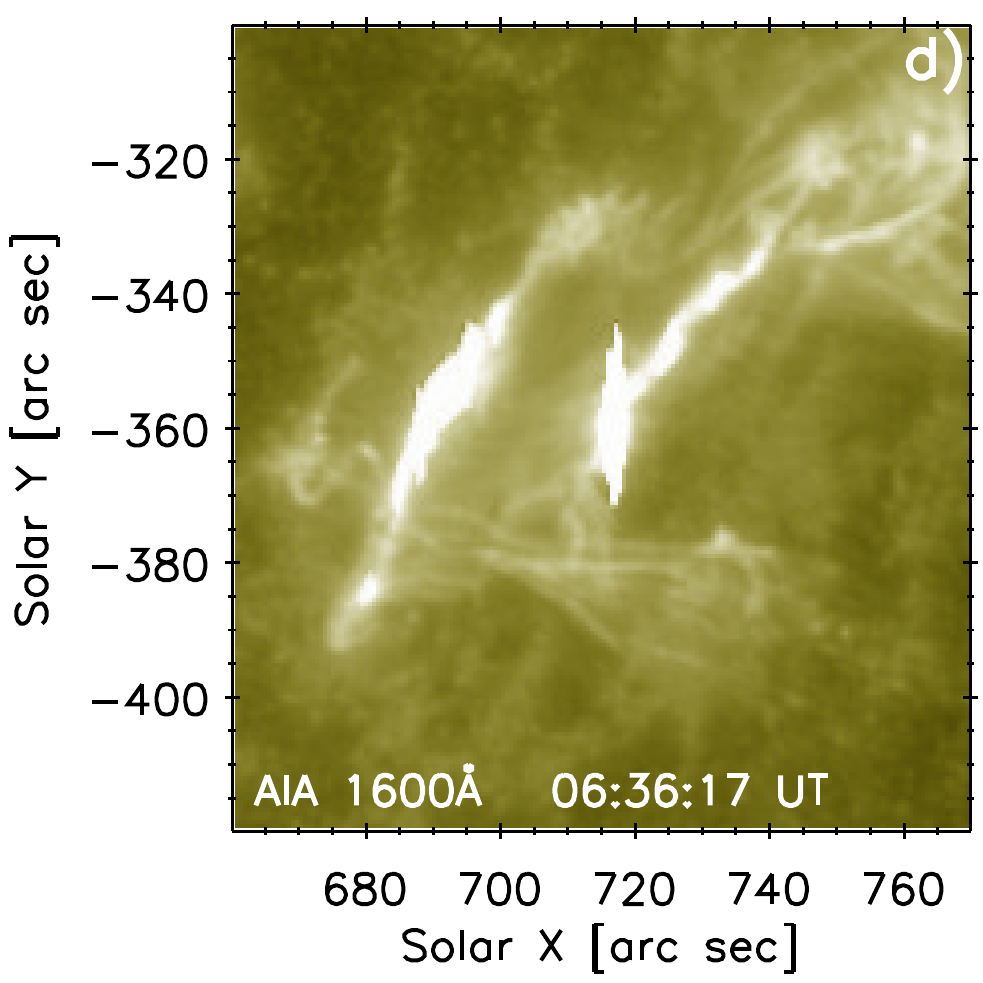}
\caption{Evolution of the flare ribbons NR and PR, as well as their hooks NRH1--2 and PRH at 06:21, 06:27, and 06:33 UT. Panel (a) shows a composite three-color image of these three times, using the blue, green, and red color, respectively. Panels (b)--(d) show the evolution as observed in AIA 1600\,\AA.
\\ (An animation of this Figure is available.)}
\label{Fig:aia1600}
\end{figure*}
\begin{figure}
	\centering
	\includegraphics[width=8.00cm,clip]{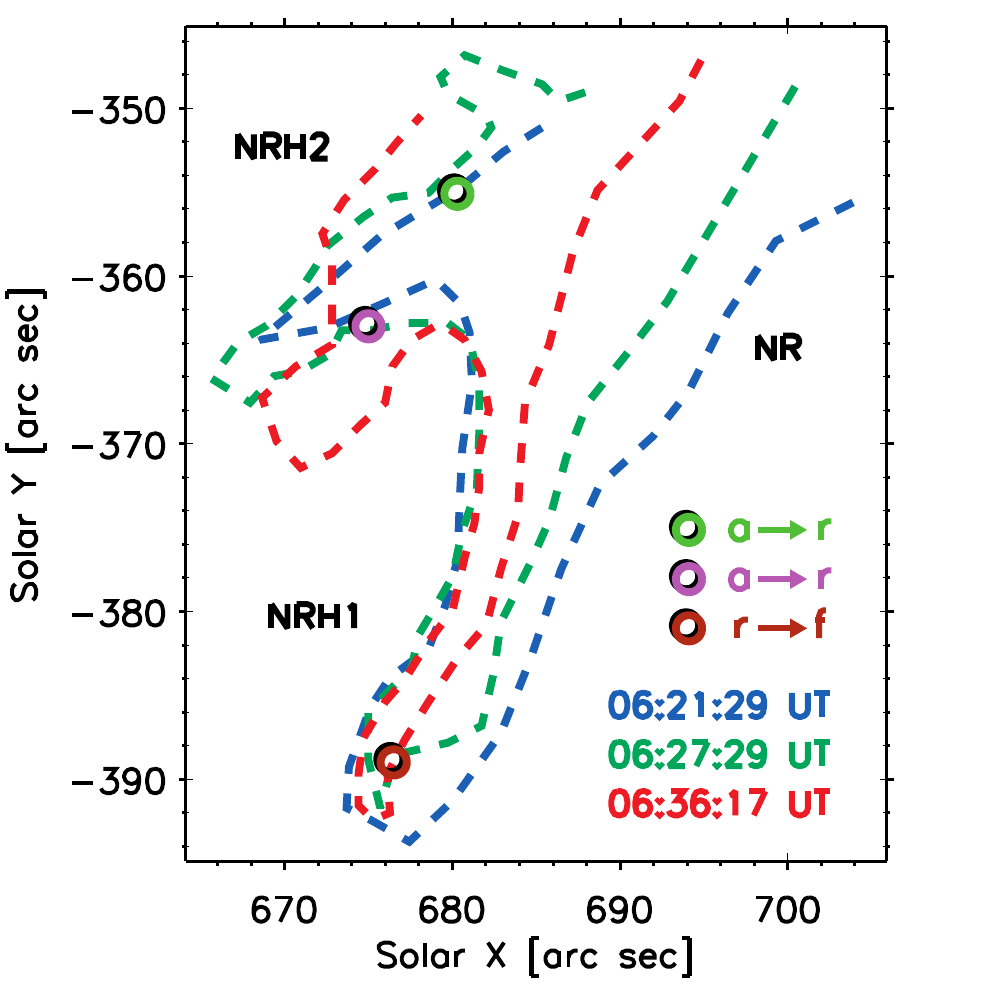}
\caption{Zoomed-in cartoon showing manual tracing of the NR and NRH1--2 at 06:21, 06:27, and 06:33 UT. Colored circles denote the locations shown in Figures \ref{Fig:aia131} and \ref{Fig:ar-rf}.}
\label{Fig:aia1600_cartoon}
\end{figure}
\begin{figure*}
	\centering
	\includegraphics[width=5.32cm,viewport= 0  0 282 278,clip]{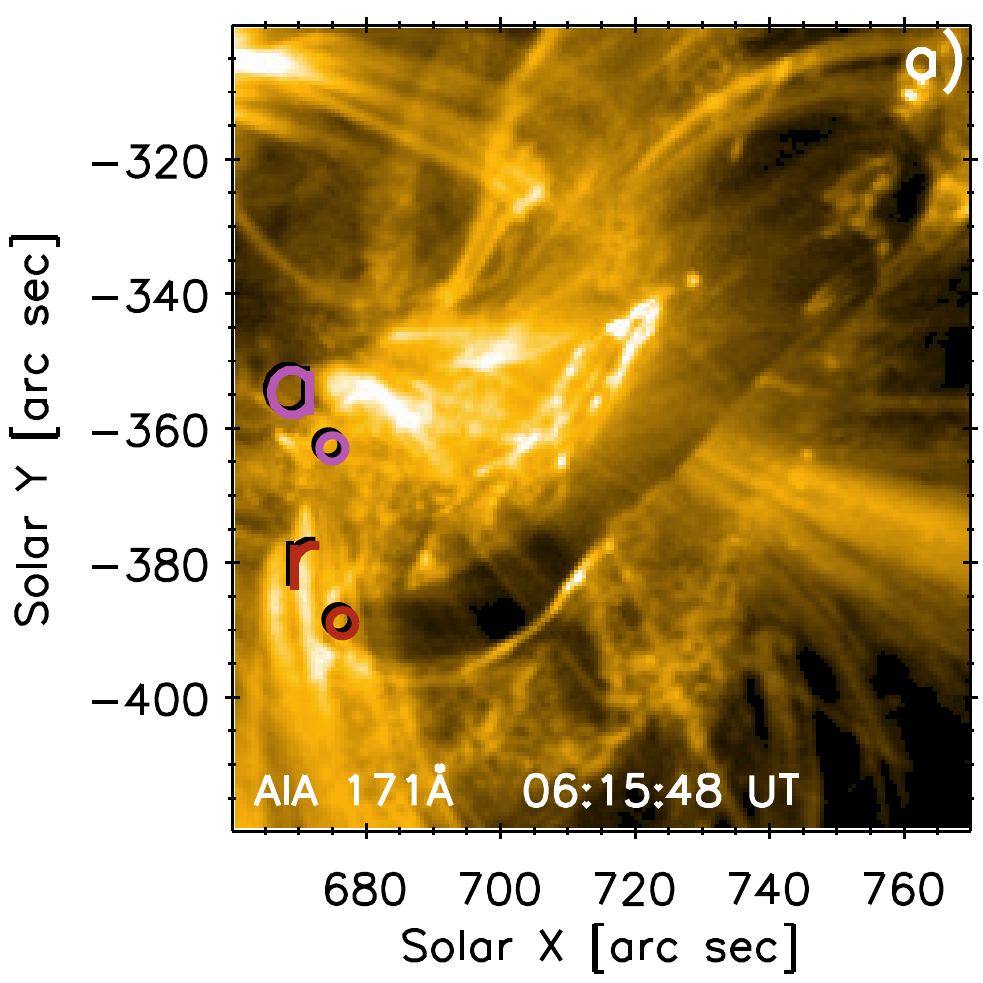}
	\includegraphics[width=4.10cm,viewport=65  0 282 278,clip]{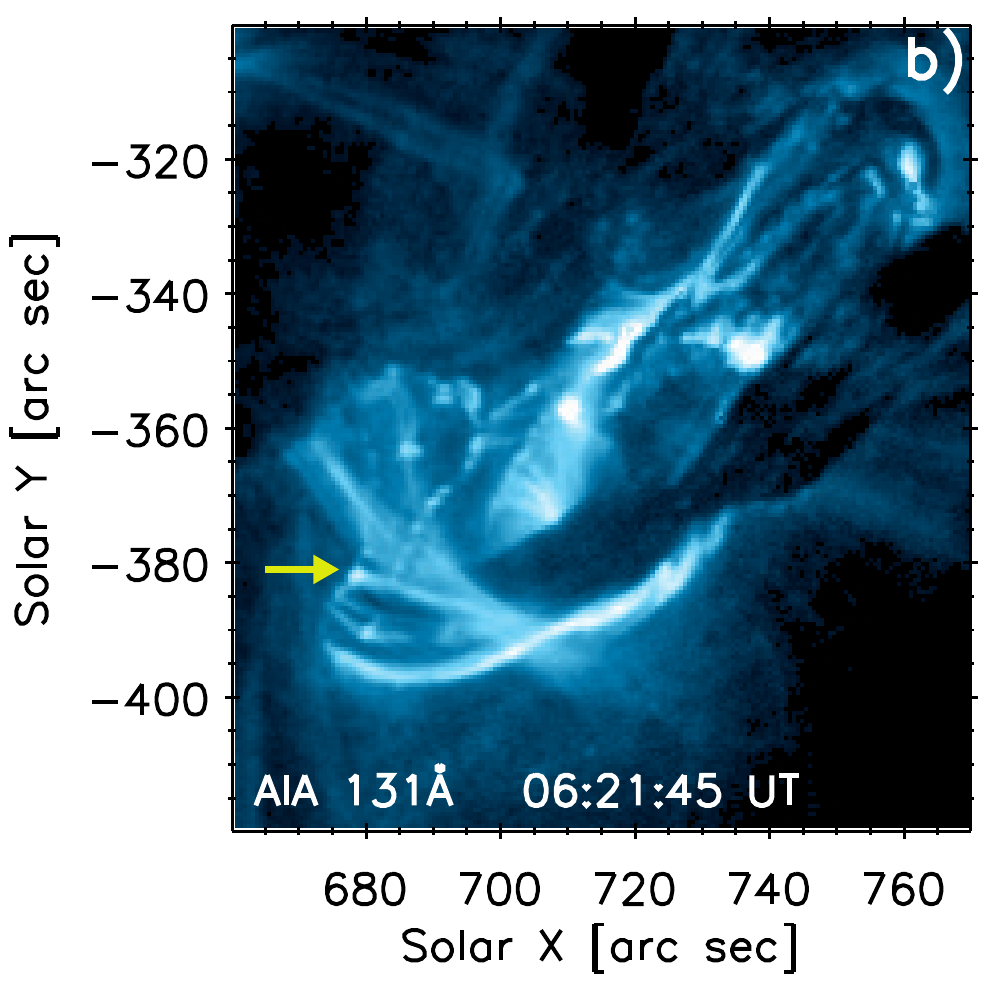}
	\includegraphics[width=4.10cm,viewport=65  0 282 278,clip]{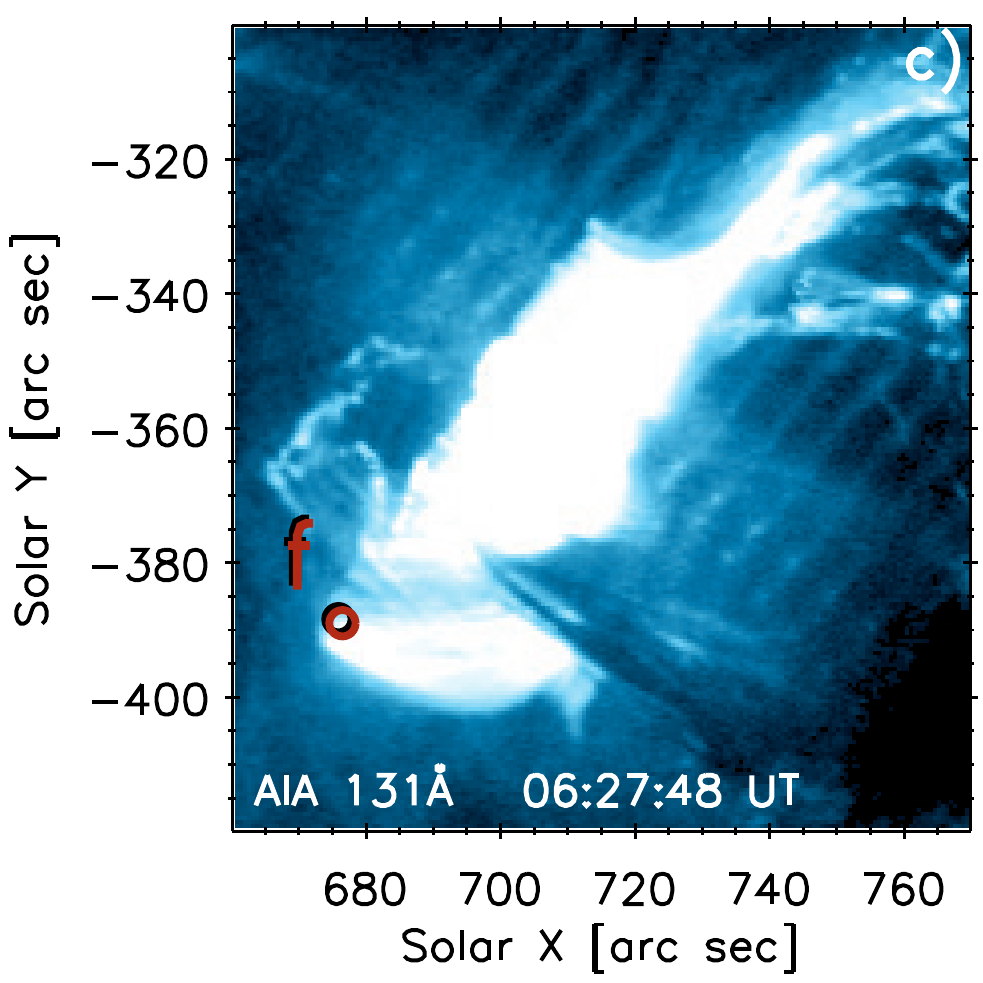}
	\includegraphics[width=4.10cm,viewport=65  0 282 278,clip]{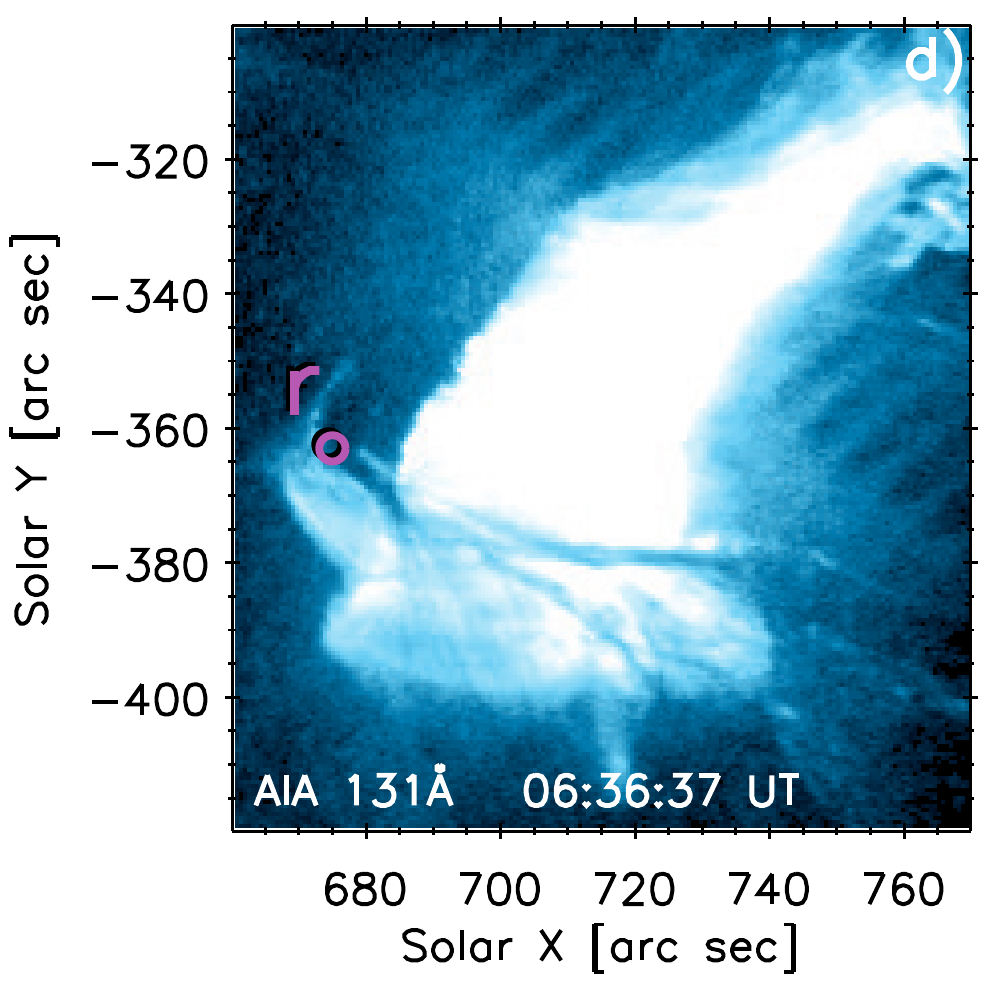}
\caption{Evolution of the filament and the hot flare emission as observed by AIA 131\,\AA. The FOV is the same as in Figure \ref{Fig:aia1600}. The colored circles denote footpoints of individual reconnecting structures. The label 'a' stands for arcade, 'r' for flux rope, and 'f' for flare loop. Yellow arrow in panel b depict flare loops rooted at the outer edge of NRH1.
\\ (An animation of this Figure is available.)}
\label{Fig:aia131}
\end{figure*}

\begin{figure*}[ht]
	\centering
	\includegraphics[width=5.41cm,viewport= 0 40 280 280,clip]{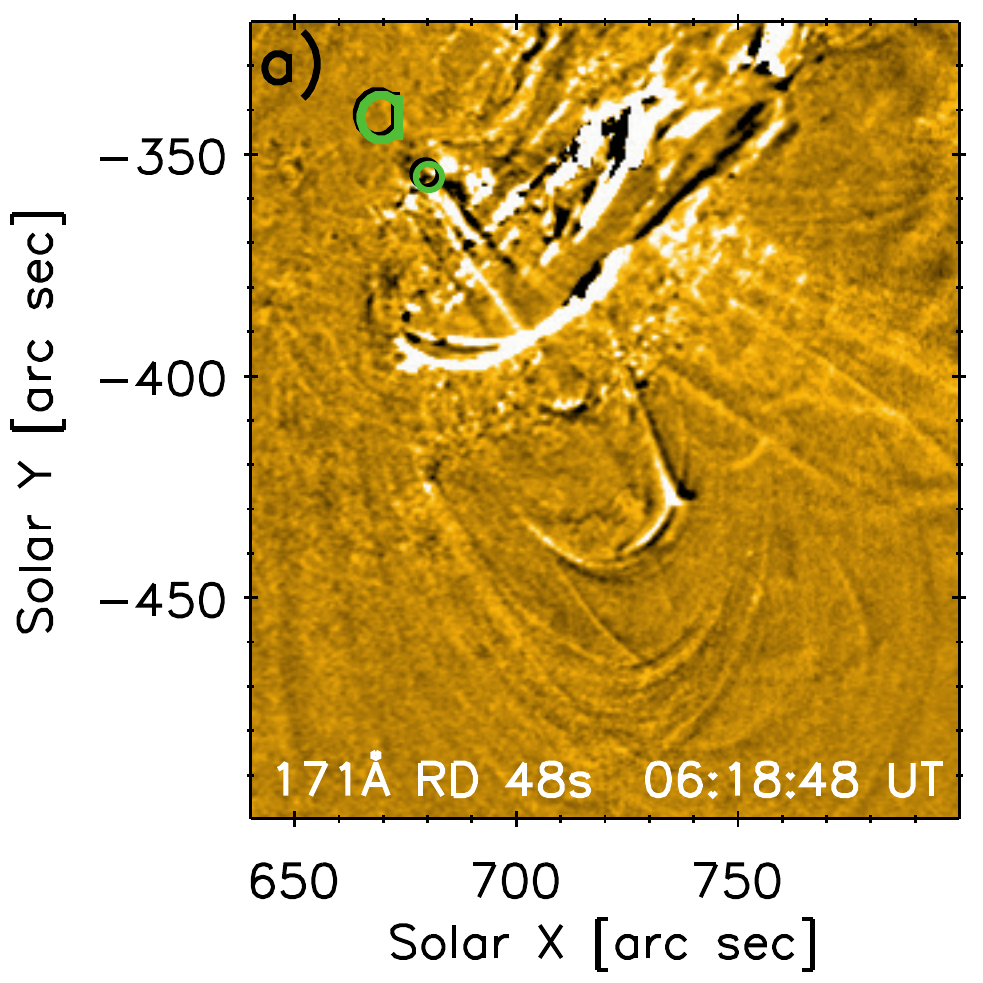}
	\includegraphics[width=4.06cm,viewport=70 40 280 280,clip]{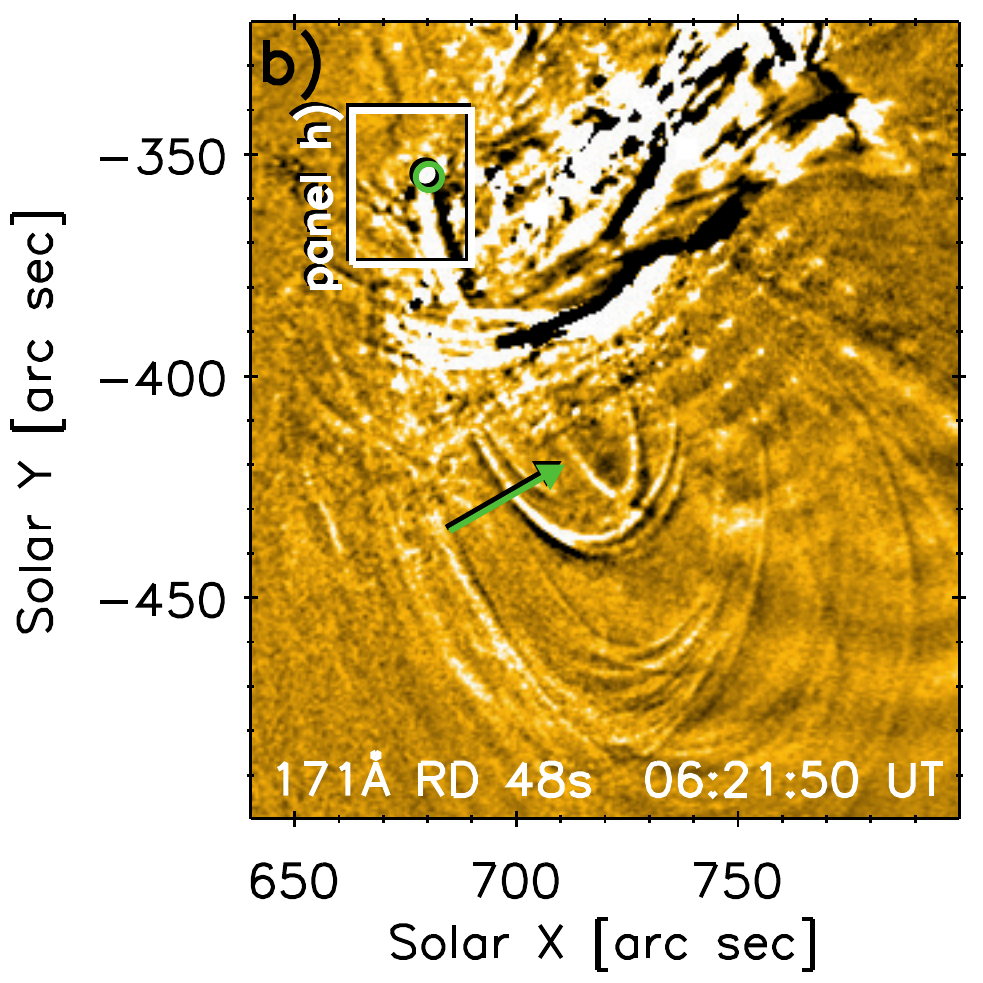}
	\includegraphics[width=4.06cm,viewport=70 40 280 280,clip]{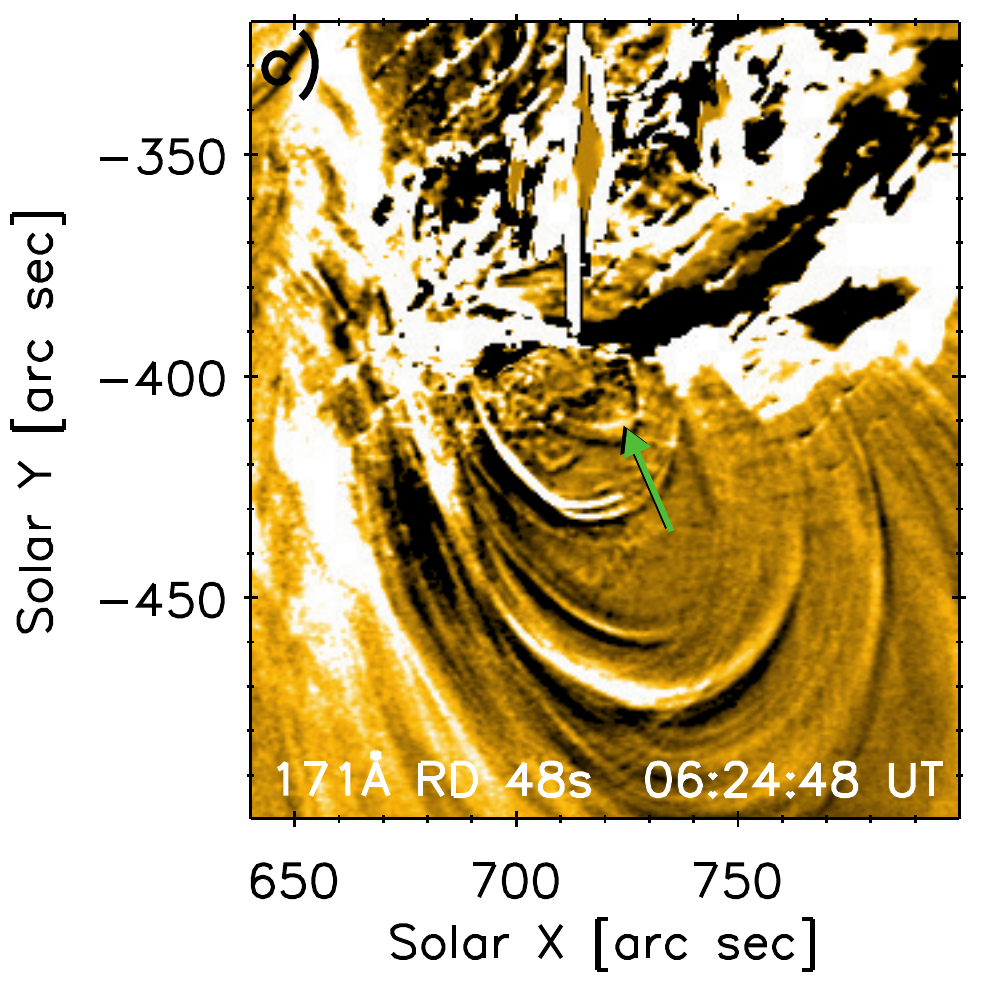}
	\includegraphics[width=4.06cm,viewport=70 40 280 280,clip]{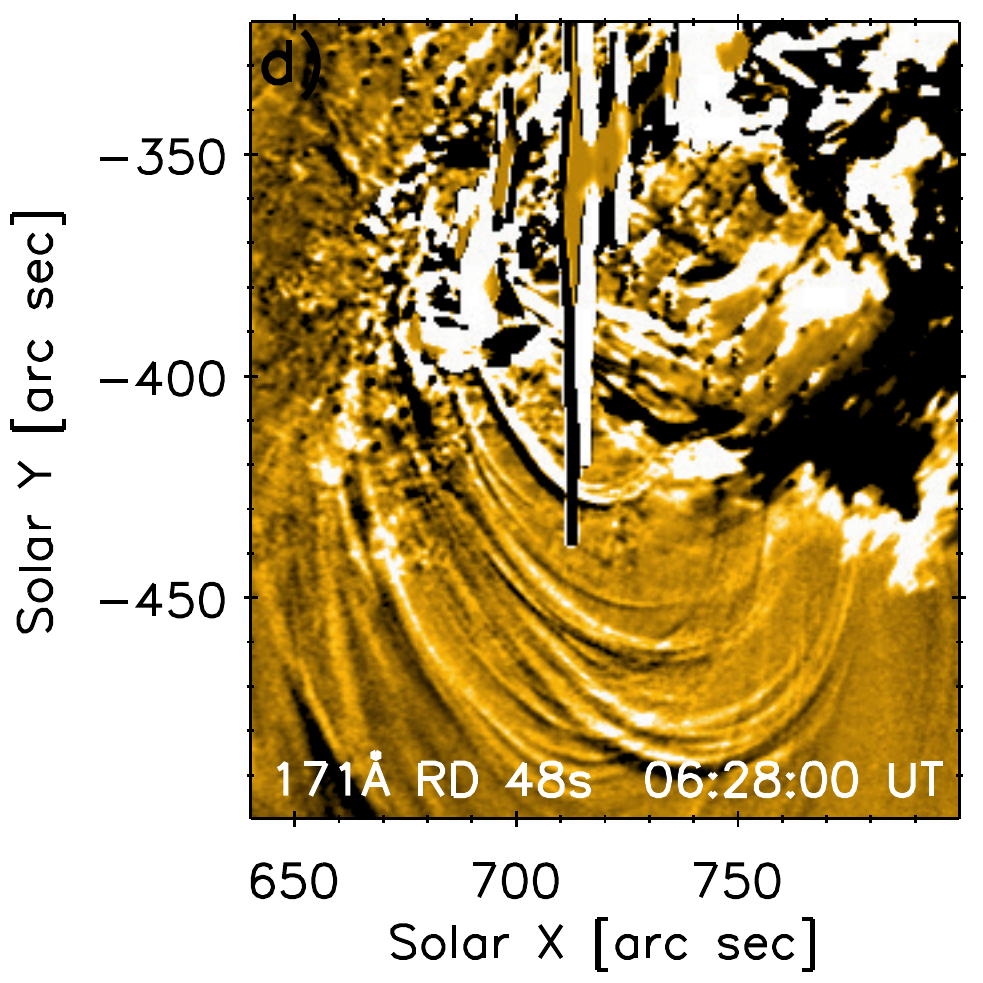}
	\includegraphics[width=5.41cm,viewport= 0  0 280 280,clip]{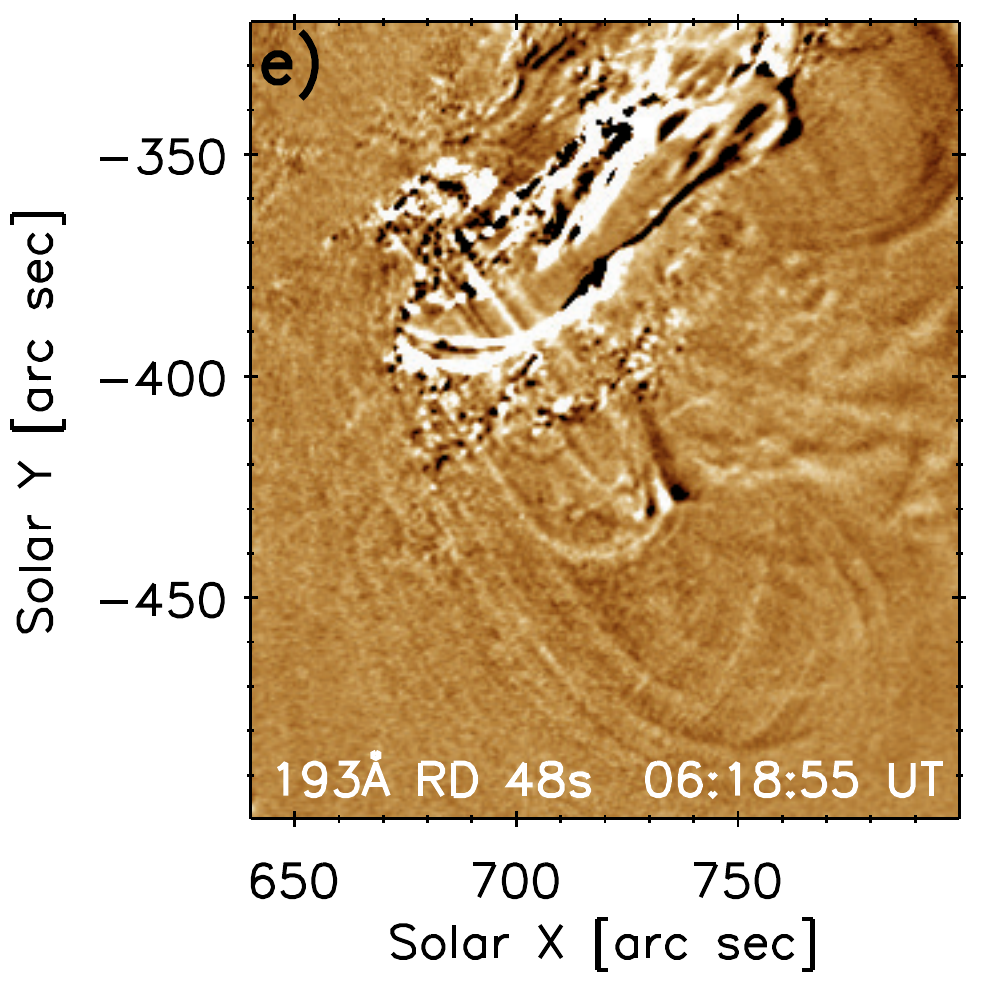}
	\includegraphics[width=4.06cm,viewport=70  0 280 280,clip]{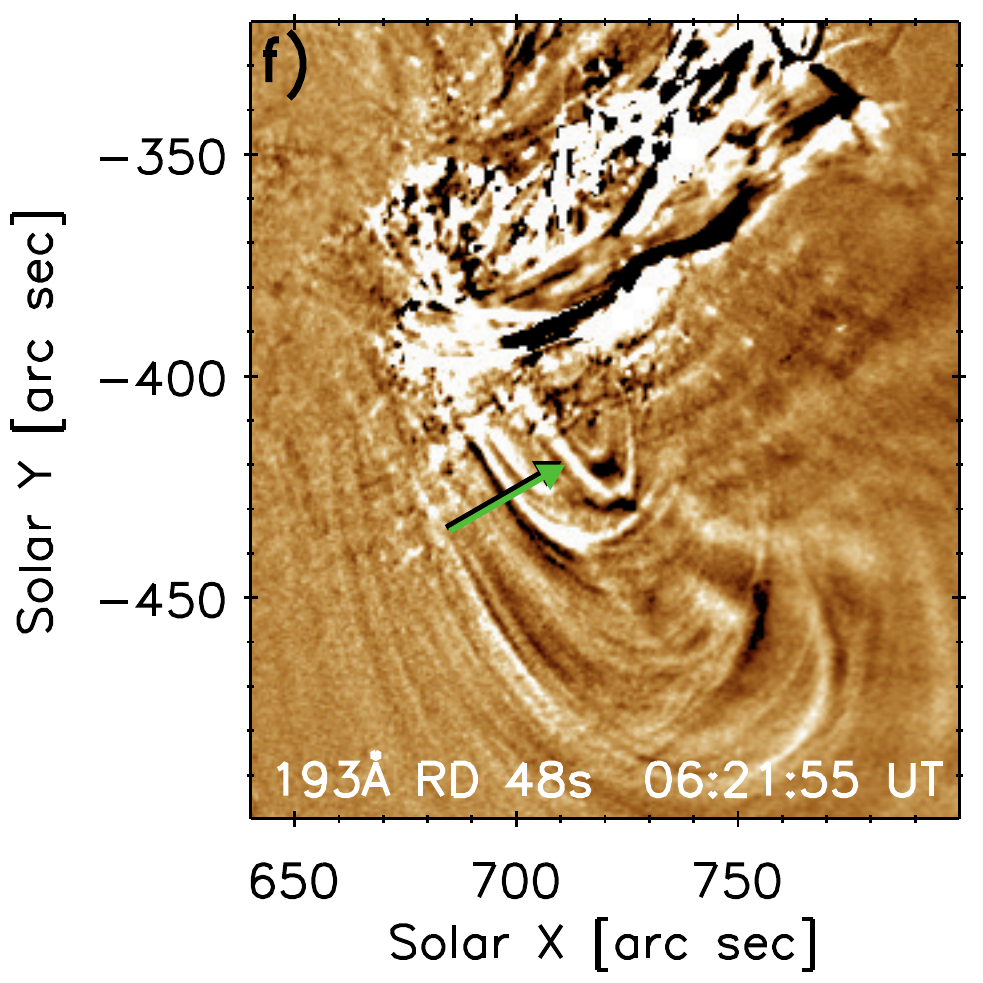}
	\includegraphics[width=4.06cm,viewport=70  0 280 280,clip]{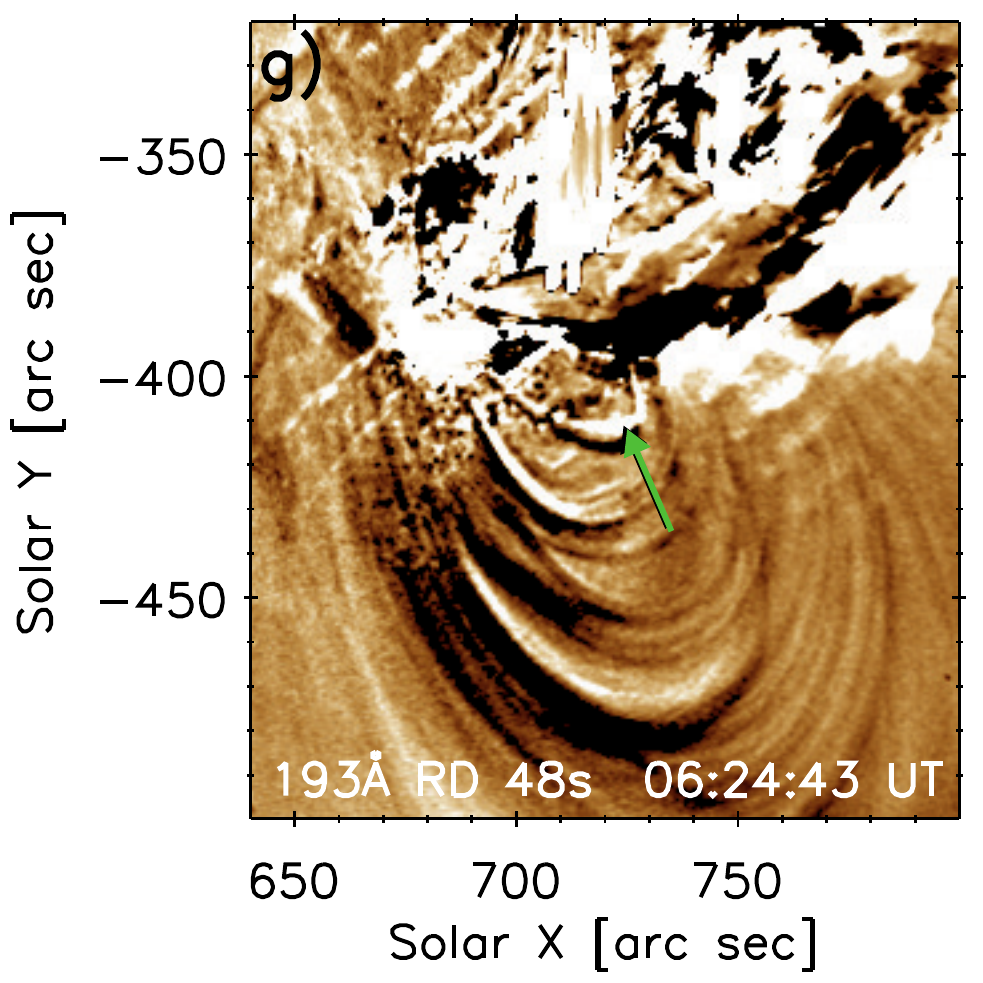}
	\includegraphics[width=4.06cm,viewport=70  0 280 280,clip]{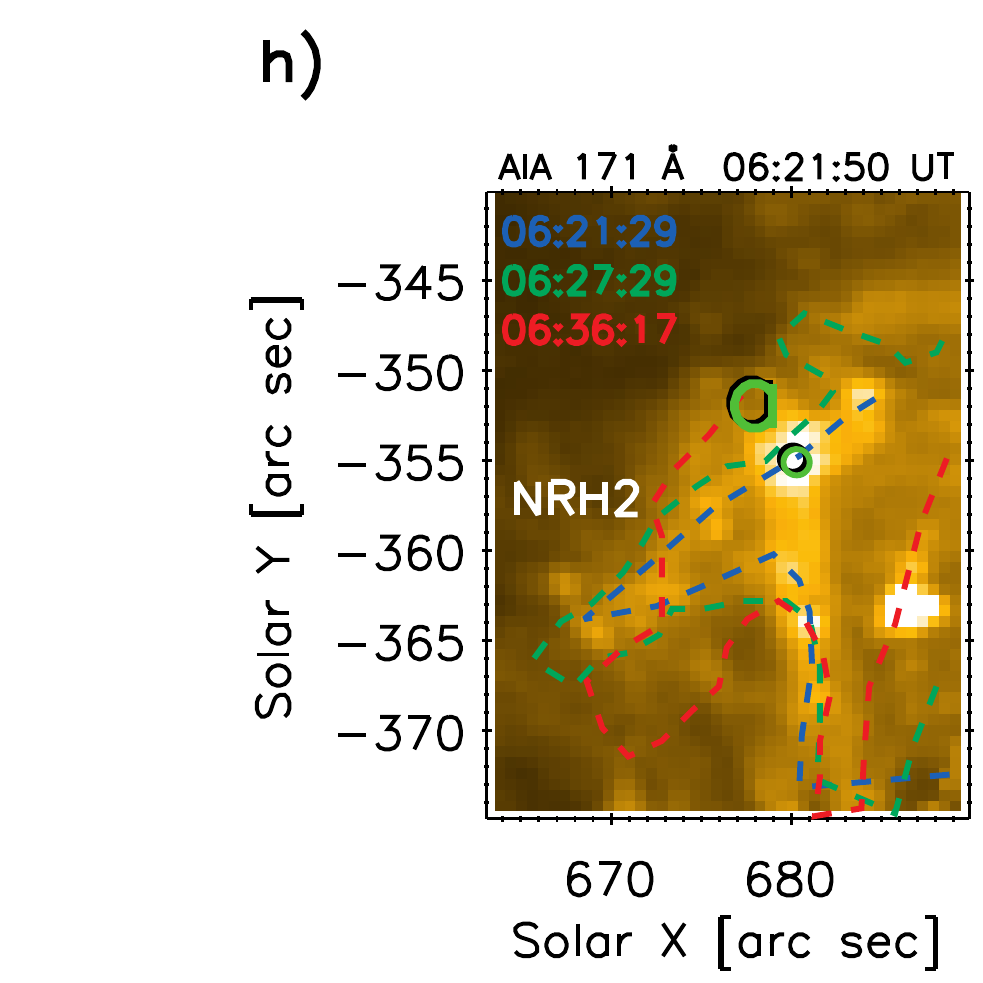}
\caption{Evolution of the overlying corona as seen in the 171\,\AA~and 193\,\AA~RD images (a--g). The color scale is saturated to $\pm$150\,DN\,s$^{-1}$, with light (dark) colors indicating increase (decrease) of emission. Green arrows denote a reconnecting coronal loop, whose footpoint is shown by the green circle denoted 'a' (see Section \ref{Sect:3.3}). Panel h shows the zoomed-in original 171\,\AA~image at 06:21:50\,UT. The location of the NRH2 at three times, corresponding to Figure \ref{Fig:aia1600}a, are overlaid by colored lines.
\\ (An animation of this Figure is available.)}
\label{Fig:ar-rf}
\end{figure*}
%
%
\begin{figure*}[ht]
	\centering
	\includegraphics[width=6.60cm,clip]{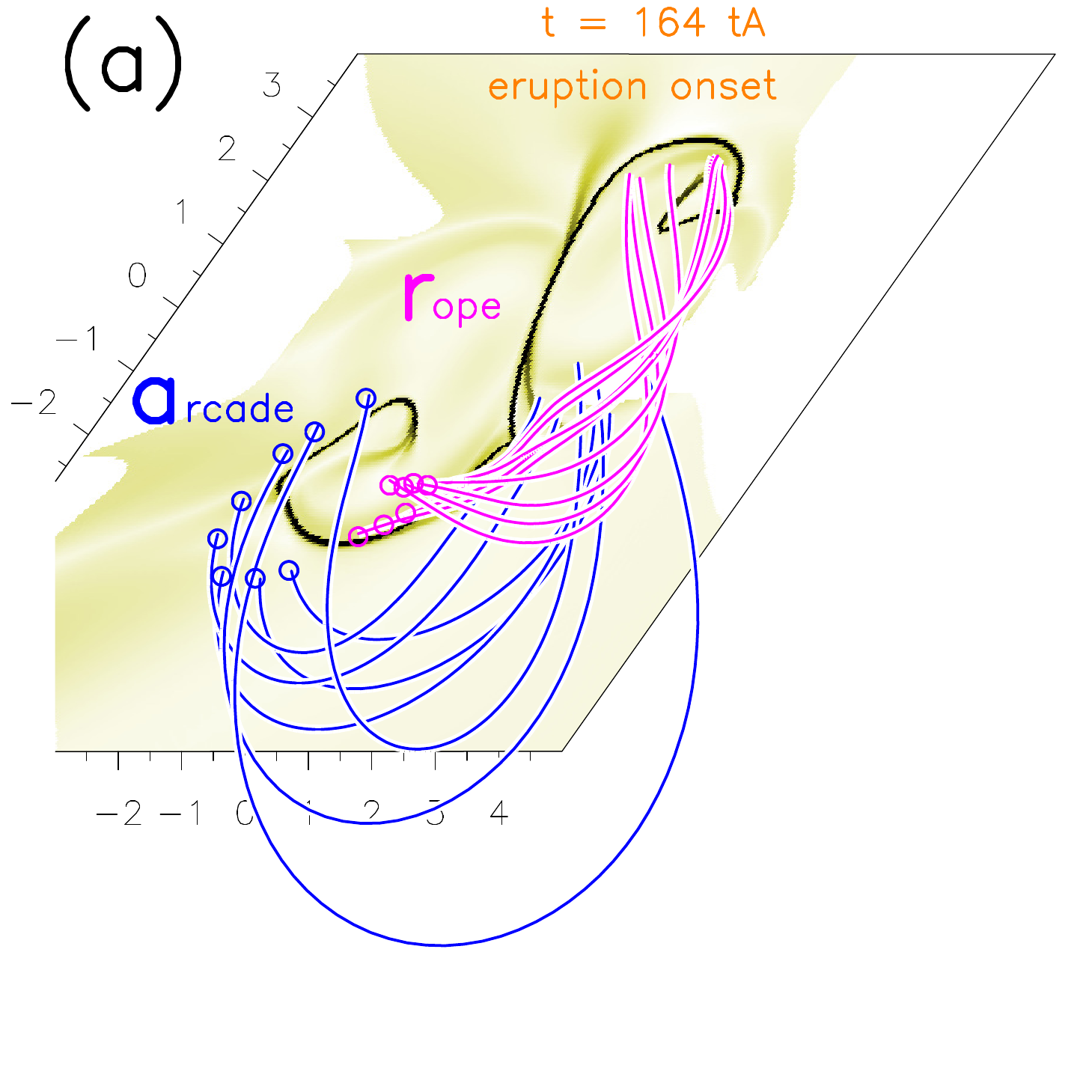}
	\includegraphics[width=6.60cm,clip]{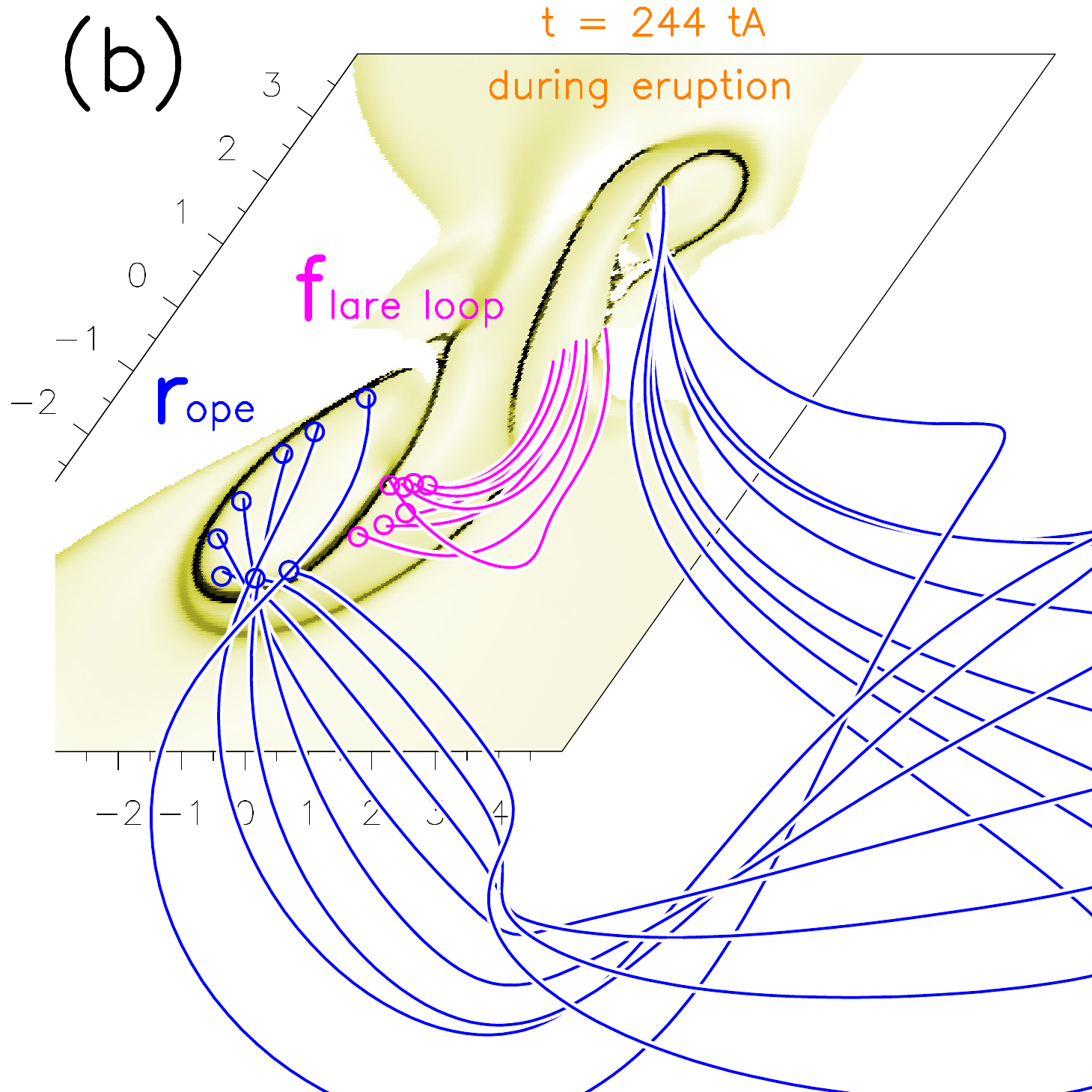}
\caption{Three-dimensional ar--rf reconnection in the 3D extensions to the standard solar flare model. The MHD simulation is the same as analyzed in \citet{Aulanier19}. The QSL footprints in the model (where the squashing factor $Q(z=0)$ reaches its maximum values), corresponding to flare ribbons, are shown by black. Selected field lines from the model are shown during (a) at the onset of the eruption, and (b) during the eruption. The footpoints near the hook are fixed and denoted by circles.}
\label{Fig:ar-rf_model}
\end{figure*}

\begin{figure*}[ht]
	\centering
	\includegraphics[width=6.61cm,viewport= 0 40 355 314,clip]{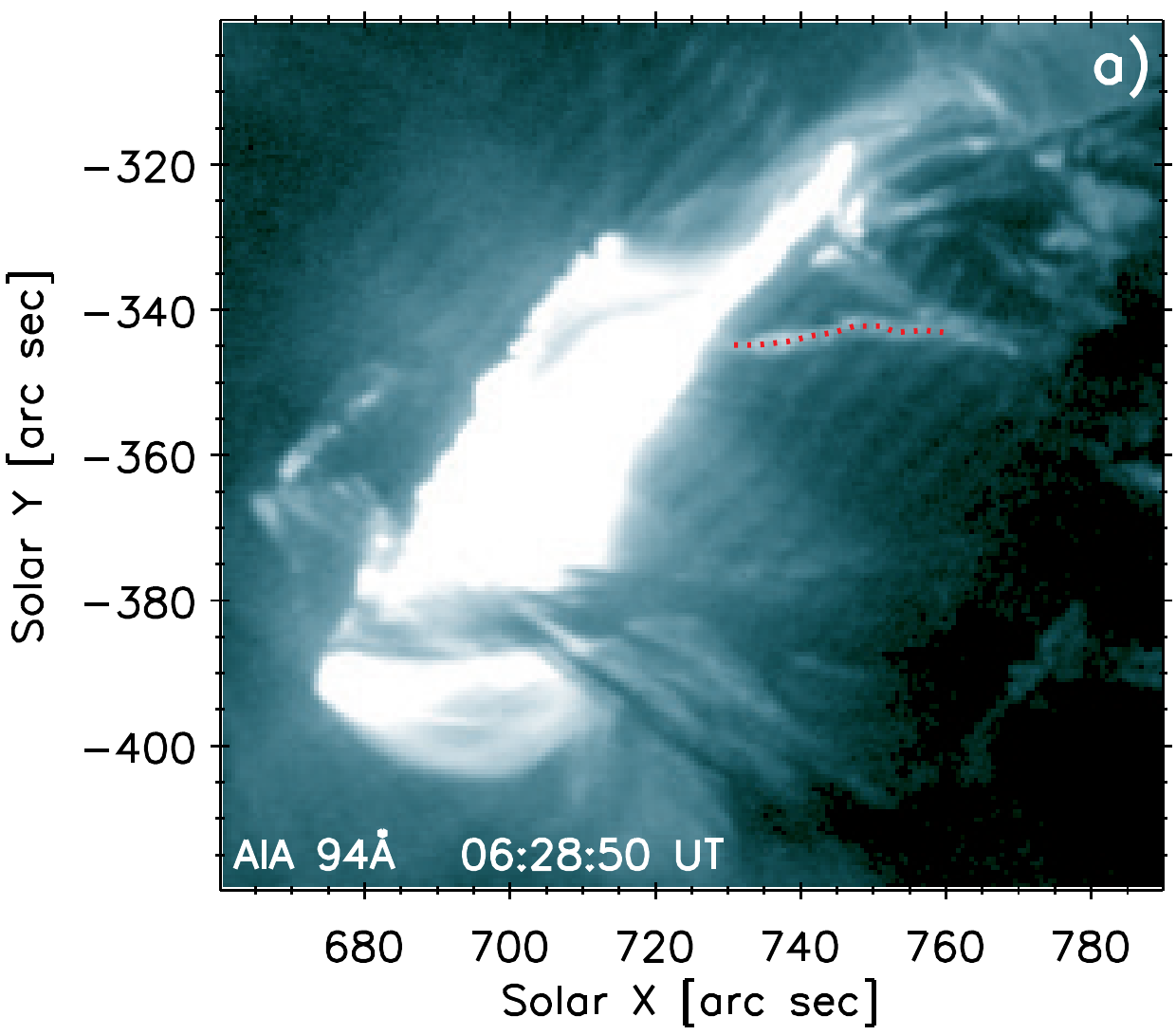}
	\includegraphics[width=5.49cm,viewport=60 40 355 314,clip]{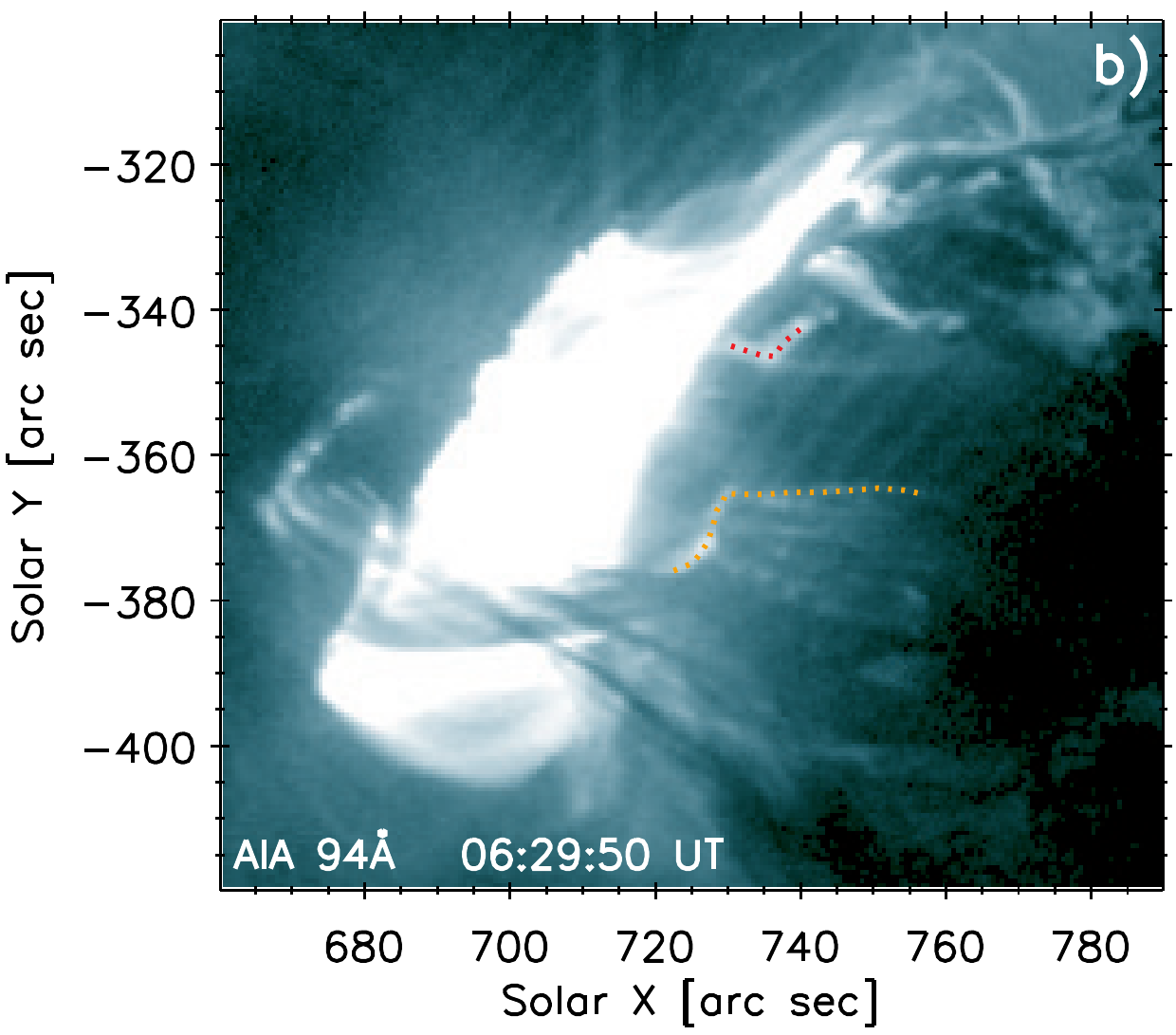}
	\includegraphics[width=5.49cm,viewport=60 40 355 314,clip]{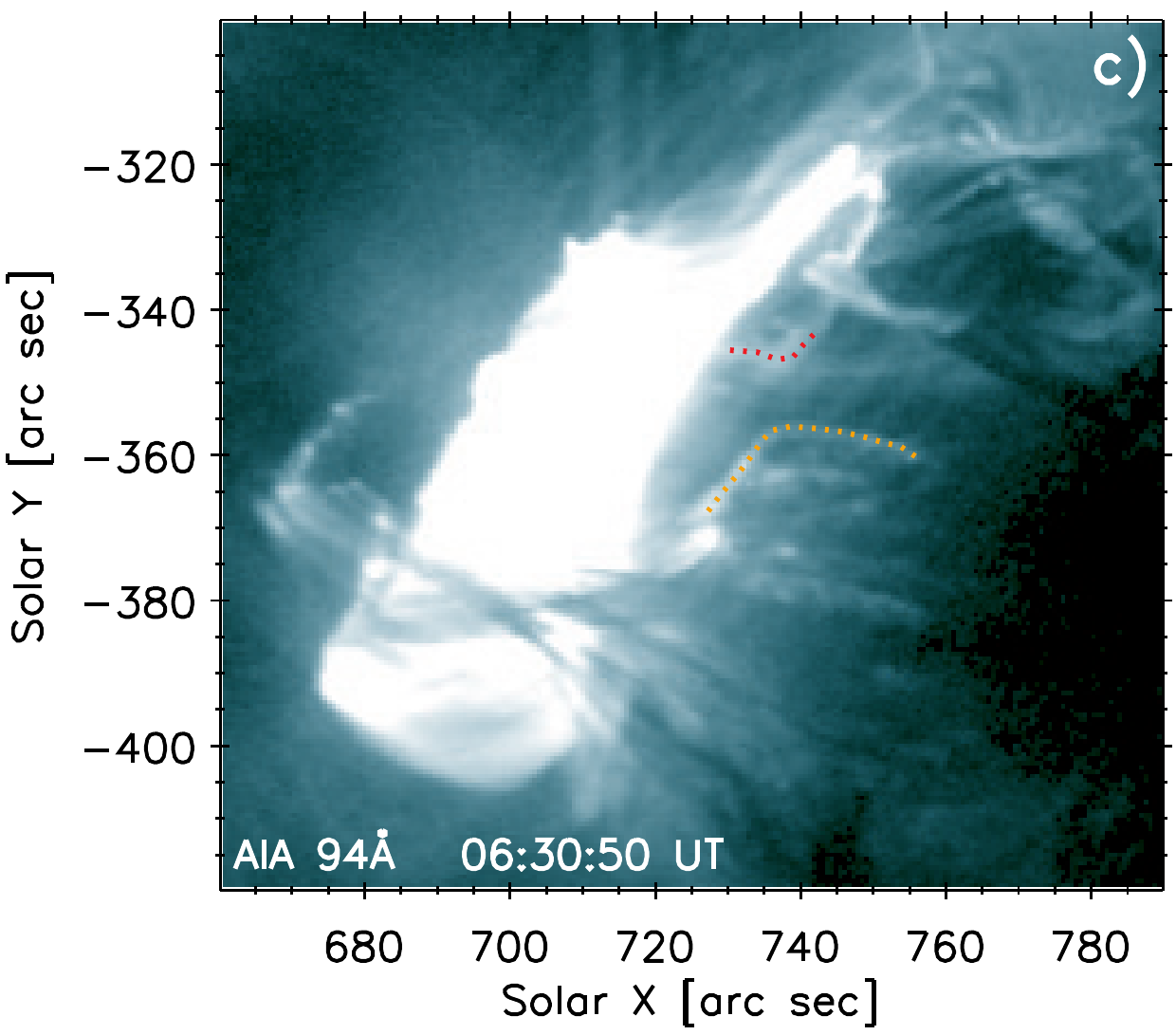}

	\includegraphics[width=6.61cm,viewport= 0  0 355 314,clip]{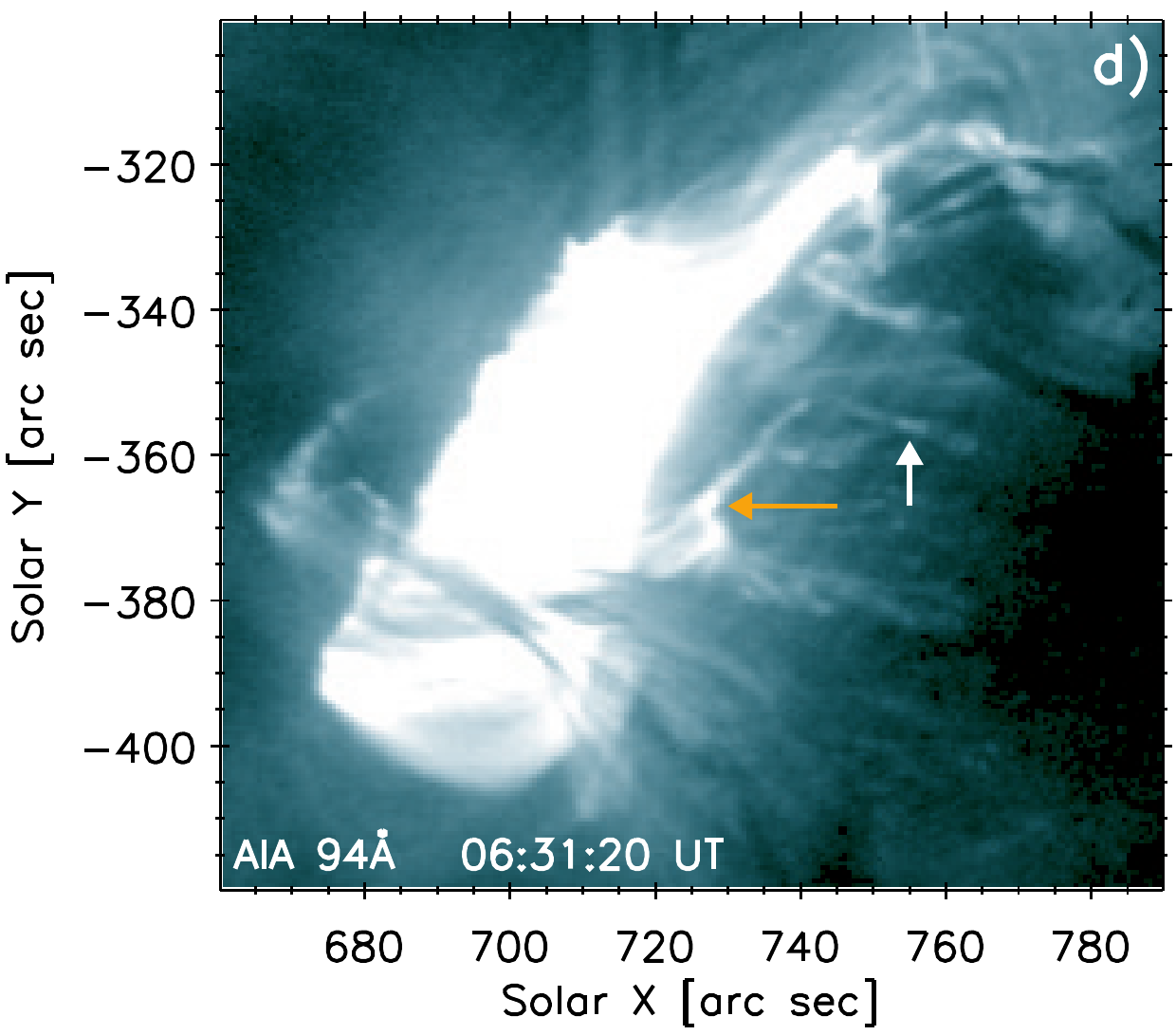}
	\includegraphics[width=5.49cm,viewport=60  0 355 314,clip]{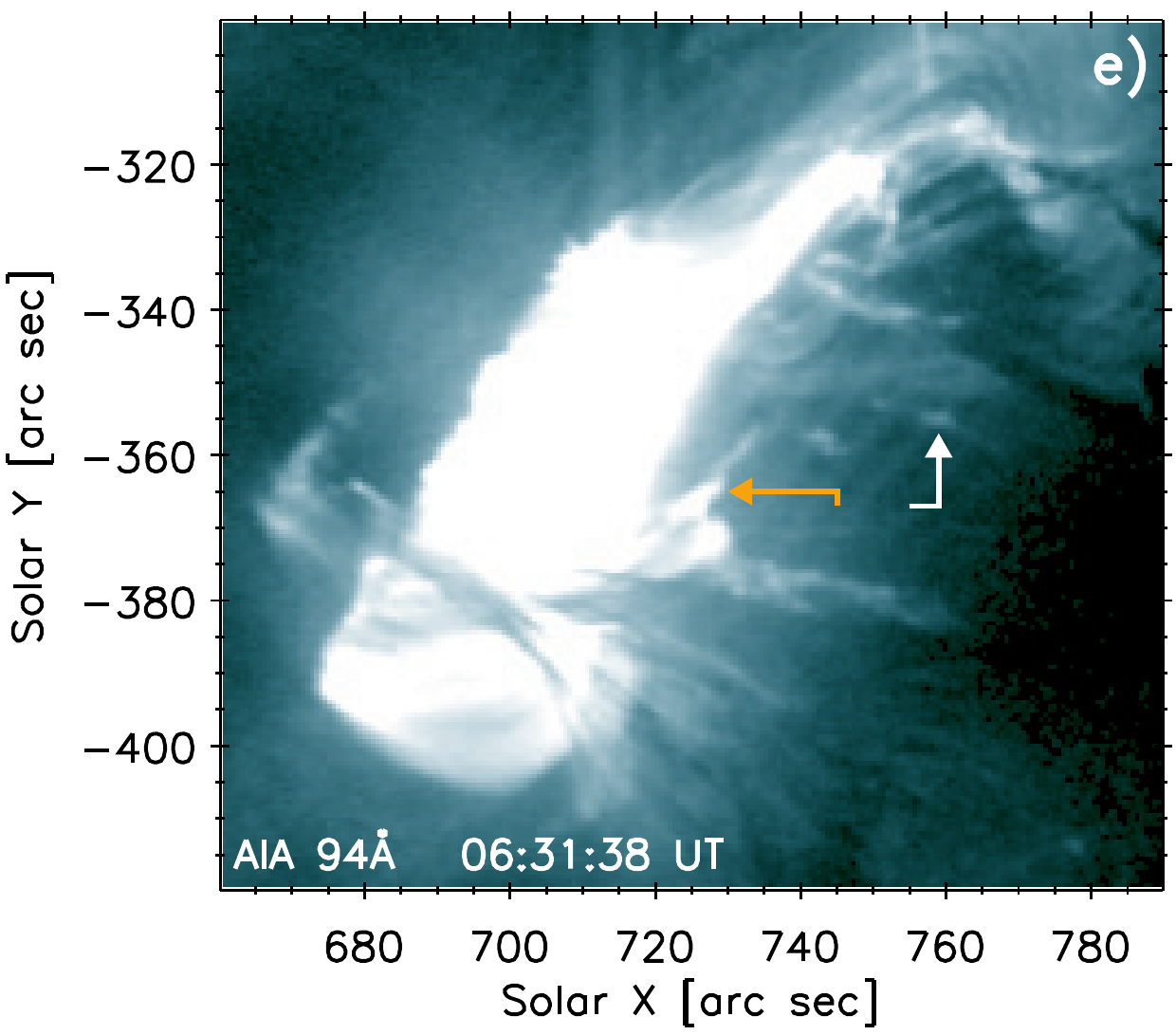}
	\includegraphics[width=5.49cm,viewport=60  0 355 314,clip]{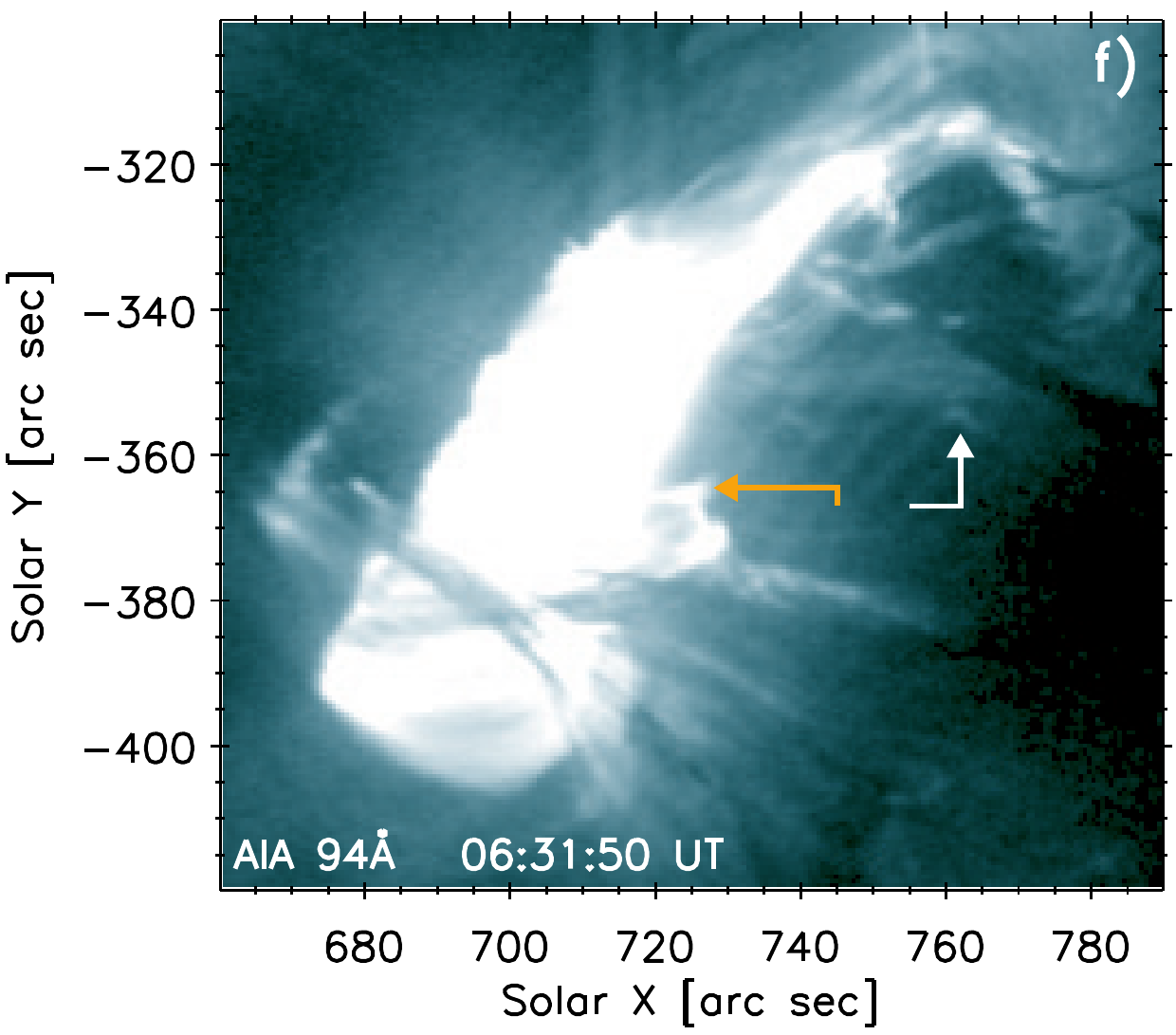}
\caption{Reconnection between bright filament strands, denoted by dotted red and orange lines, during the peak phase as observed by AIA 94\,\AA. Orange arrows show the newly-formed flare loop, while the white arrow denote a bright blob ascending into the erupting filament. The starting points of the arrows are kept the same.
\\ (An animation of this Figure is available.)}
\label{Fig:rr-rf}
\end{figure*}

%
\section{The Filament Eruption of 2011 June 07}
\label{Sect:2}

The filament eruption studied here occurred within the active region NOAA 11226 (hereafter, AR 11226), see also Figure \ref{Fig:Eruption}c). The AR is the westernmost part of a larger complex of ARs, including AR 11227 and small AR 11233 between them. The eruption involved an unusually massive filament \citep{vanDriel14} which became the bright core of a 3-part CME \citep{Howard17}. It was associated with a long-duration M2.5--class flare that started at $\approx$06:08\,UT, reached the impulsive phase at 06:27\,UT, and exhibited a broad peak around 06:41\,UT (Figure \ref{Fig:Eruption}a). The Solar Object Locator for this event is SOL2011-06-07T06:24:00.

The eruption is a well-known event. \citet{Inglis13} studied the motion of hard X-ray emission within the flare ribbons. \citet{vanDriel14} studied the reconnections between the erupting filament and the neighboring ARs within the AR complex, occurring well after the peak phase of the flare. \citet{Yardley16} investigated the evolution of the magnetic field and found that strong flux cancellations contributed to the buildup of this large filament. Properties of filament blobs, some of which were falling back to the Sun, were investigated by \citet{Innes12}, \citet{Williams13}, \citet{Carlyle14}, and \citet{Thompson16}; the impacts themselves by \citet{Gilbert13}, \citet{Reale14}, and \citet{Innes16}. Bright filament ejecta in the outer corona were studied by \citet{Susino15}, \citet{Wood16}, and \citet{Mishra18}. However, none of these studies dealt with the initial and impulsive phases of the eruption, and the accompanying reconnection in the low corona at these times.

To do that, we use the imaging data from the Atmospheric Imaging Assembly \citep[AIA][]{Lemen12} onboard the \textit{Solar Dynamics Observatory} \citep[SDO][]{Pesnell12}. AIA provides full-Sun observations in 10 spectral channels covering wide range of temperatures, from the photosphere to the corona. Its spatial resolution is 1$\farcs$5 with a pixel size of 0$\farcs$6. Hot flare emission is documented in the 94\,\AA~and 131\,\AA~channels \citep{ODwyer10}. The coronal environment is observed in 171\,\AA, 193\,\AA, 211\,\AA, and 335\,\AA. In these filters, the filament is visible as a dark structure due to the absorption by H and He continua \citep[e.g.,][]{Heinzel08,Williams13}. The flare ribbons are observed in the 1600\,\AA~channel containing \ion{C}{4} emission \citep{Lemen12,Simoes19} originating in the transition region.

The filament eruption in AIA 193\,\AA~is shown in Figure \ref{Fig:Eruption}. At 06:15\,UT (panel b), the filament is already rising. Ribbon brightenings can be discerned near its southern leg at [$X$,$Y$]\,$\approx$\,[675$\arcsec$,$-390\arcsec$]. This filament leg is overlaid by a multitude of closed coronal loops anchored in the vicinity. Towards the impulsive phase, at 06:25\,UT, the filament undergoes a full eruption. Prominent flare ribbons are observed nearby (panels d--e). The 1600\,\AA~image together with the \textit{SDO}/HMI (Helioseismic and Magnetic Imager) line-of-sight magnetogram \citep{Scherrer12} shows that the southern leg of the filament is located in the negative polarity, and is encircled by a hook NRH1 of the negative ribbon NR. In addition, a second hook NRH2 is present, located further north (Figure \ref{Fig:Eruption}e). In reality, both NRH1 and NRH2 are likely a part of a single hook, deformed by the presence of other active regions east of it \citep[see Figure 1c and Figures 9b,d; 10a of][]{vanDriel14}. We will nevertheless refer to the observed morphology and use the labels NRH1 and NRH2. The conjugate ribbon PR and its hook PRH are also discernible.

During the eruption, many of the filament strands have brightened, and are visible both in 193\,\AA~and 1600\,\AA~(panels d--e). Near the peak of the flare at 06:35\,UT, the filament is already escaping the Sun. Its darkening is probably a combination of decreased AIA 193\,\AA~exposure time and removal of foreground emission during the eruption. The growing arcade of bright flare loops is evident in 193\,\AA~(panel f). PR and NR spread away, with the straight NR closing NRH1~(panel g). The closing of NRH1 is accompanied by a \textit{drift} of the southern leg of the filament to NRH2. At 06:35\,UT, it is at position [$X$,$Y$]\,$\approx$\,[675$\arcsec$,$-365\arcsec$], a shift of about 25$\arcsec$ northward. Note that this happens well before the reconnection of the erupting filament with surrounding ARs, which occurs only at $\approx$07:00\,UT \citep{vanDriel14}.

%
%
\section{AR--RF Reconnection during the Eruption}
\label{Sect:3}

To test whether the observed drift of the filament leg could be due to the ar--rf reconnection predicted by \citet{Aulanier19}, we study the AIA observations of both the flare and surrounding corona.

\subsection{Ribbon evolution}
\label{Sect:3.1}

We first examine the evolution of the ribbons and their hooks. The AIA 1600\,\AA~observations at 06:21, 06:27, and 06:36\,UT are shown in Figure \ref{Fig:aia1600}. These times cover the early, impulsive, and peak phases of the flare, respectively. The ribbons become well-visible after about  06:18\,UT. Initially, NR and PR are close together near the polarity inversion line (PIL). They subsequently separate with relative velocity of about 26\,km\,s$^{-1}$. This separation causes NR to drift into NRH1, narrowing it (Figure \ref{Fig:aia1600}a), since the outer edge of NRH1 does not move.

A composite image of the ribbon evolution at the three times of Figure \ref{Fig:aia1600}b--d is shown in panel\,a, using blue, green, and red colors for 06:21, 06:27, and 06:36\,UT, respectively. In addition, Figure \ref{Fig:aia1600_cartoon} shows the zoomed-in manual tracing of the shape of NRH1--2 at these three times. The ribbon tracing is done analogously to \citet[][Figure 7e--f therein]{Aulanier19}. We note that the tracing is only approximate due to finite width of the ribbon, varying intensity along it, and local saturation. The uncertainty is estimated to be at least the AIA resolution of $\pm$1$\farcs$5, and is larger within saturated areas.

Figures \ref{Fig:aia1600} and \ref{Fig:aia1600_cartoon} show that the evolution of the NRH2 is a complex one. It expands between 06:21--06:27\,UT (blue--green), then changes shape as its curved part (elbow) moves further west at 06:36\,UT (red; Figure \ref{Fig:aia1600}a). Similarly to the flares investigated previously by \citet{Aulanier19} and \citet{Zemanova19}, there exists an area (around $X$\,=\,$670\arcsec$, $Y$\,=\,$-365\arcsec$) which is outside of the flux rope (NRH2) at 06:21\,UT, then inside the flux rope at 06:27\,UT, but is again outside of if at 06:36\,UT. In line with previous works, this suggests probable sequences of ar--rf reconnections in the framework of the model predictions \citep{Aulanier19}.

\subsection{Evolution of filament and flare loops}
\label{Sect:3.2}

We next examined the evolution of the flare itself. Flare emission is well-observed in AIA 131\,\AA~and 94\,\AA, both of which show the same morphology of flare loops. The flare emission is detected as early as 06:08\,UT and comes primarily from the ribbon hooks. Here, we focus on NRH1 and NRH2, since PRH is largely obscured by the erupting filament. Figure \ref{Fig:aia131} shows the evolution of the flare emission in 131\,\AA. The location of the southern leg of the filament is also evident as a dark, evolving structure in this passband. Early in the flare, the leg of the rising filament in NRH1 is flanked on it southern edge by flare loops. To better identify the initial filament footpoint, the 131\,\AA~image at 06:15\,UT was replaced by the 171\,\AA~one, which does not contain the flare emission. The filament footpoint there is denoted by red circle 'r'.

At 06:21\,UT (Figure \ref{Fig:aia131}b), the evolving flare loops envelop the filament leg. Some of these flare loops are pointed out by yellow arrow. The flare loops rooted in NRH1 slip along its outer edge with apparent speeds of about 23\,$\pm$\,3\,km\,s$^{-1}$ (see animation accompanying Figure \ref{Fig:aia131}). At 06:27\,UT, the original position of the filament leg, denoted by red circle (Figure \ref{Fig:aia131}c), no longer corresponds to the filament leg, but to footpoints of flare loops 'f' within a growing flare loop arcade. The filament leg itself drifts northward: In Figure \ref{Fig:aia131}c, the filament leg is already located to the north of the red circle. We note that the drift itself is not easy to follow, for example in a time-distance plot, due to many overlapping structures, as well as saturation and diffraction from bright flare loops, and changes in AIA exposure times. However, at about 06:36\,UT, the filament footpoint clearly reached NRH2 (Figure \ref{Fig:aia131}d), more than 25$\arcsec$ northward. This shift is much larger than the AIA resolution. Furthermore, the filament leg in NRH2 is not overlapped by the bright flare arcade, permitting its unambiguous identification. In Figure \ref{Fig:aia131}d, the new position of the filament leg inside NRH2 is marked by violet circle 'r'.

Inspecting the surrounding coronal emission, we find that before the flare, at 06:15\,UT, the location of the violet circle corresponded to a footpoint of a faint coronal loop seen in 171\,\AA, denoted again by a violet circle and letter 'a'. Thus, for the first time, we identify all four constituents of the 3D ar--rf reconnection: The original rope 'r' reconnects with a coronal arcade 'a' to become a flare loop 'f' and a new rope field line 'r'. The temporal gap between the occurrence of the flare loop (06:27\,UT) and the new position of the filament leg in NRH2 (06:36\,UT) could be explained by the time the filament material needs to fall back from high altitudes where it was carried during the eruption. Note that \citet{vanDriel14} shown that falling filament material can fall down at locations initially unconnected to the eruption by means of reconnections with loops from neighboring active regions. Here, the mechanism is analogous, but occurs earlier during the eruption, and within the same erupting active region.

The locations of the red and violet circles, where the footpoints of the constituents of the ar--rf reconnection are rooted, are also shown in Figure \ref{Fig:aia1600}a with respect to NRH1 and NRH2. It is seen that the red circle is swept by the outward motion of NR away from the PIL, in agreement with this location becoming a flare loop. Meanwhile, the violet circle enters the growing NRH2 and stays inside, in agreement with this location becoming a footpoint of a flux rope field line \citep[][and references therein]{Aulanier19}. We note that in this flare, the time the ribbon sweeps a particular feature can be difficult to identify, owing to the finite AIA resolution of 1$\farcs$5. This uncertainty is translated into uncertainties of identification of both the footpoint locations as well as the ribbon.


%
\subsection{Reconnection of overlying coronal loops}
\label{Sect:3.3}

Although in Section \ref{Sect:3.2} we identify the four constituents of the ar--rf reconnection, there is a multitude of coronal loops overlying the filament, some of which may also reconnect with the filament. Identifying such reconnecting loops is however not an easy task, since the coronal loops can be faint in 171\,\AA, 193\,\AA, and 211\,\AA~(hundreds of DN\,s$^{-1}$) compared to the background and especially the flare. The overall evolution of coronal loops is nevertheless visible in the running-difference (hereafter, ``RD'') images, constructed with a time delay of 48\,s. We chose such delay to avoid the AIA exposure control affecting every second exposure, and to enhance motion of structures compared to a 24\,s difference. 

A series of 171\,\AA~and 193\,\AA~RD images is shown in Figure \ref{Fig:ar-rf}. These RD images are saturated to $\pm$150\,DN\,s$^{-1}$, with the conventional color scaling, where light and dark colors indicate increase and decrease of emission, respectively. It is readily seen that there are many overlying loops which exhibit motions during the eruption. However, not all of the loops reconnect with the filament: Some constitute the compression front (visible at 06:24\,UT at $X$\,$\approx$\,650$\arcsec$; Figure \ref{Fig:ar-rf}c). Nevertheless, a series of reconnection-candidate loops rooted near NRH2 can be identified. Due to many overlapping structures, the identification of their footpoints is not straightforward. One prominent example loop is however pointed out by green arrow, while its footpoint is denoted by green circle. At 06:21\,UT, the upper portion of this loop is better visible in the 193\,\AA~RD image, while its lower leg is more intense in 171\,\AA~RD (Figure \ref{Fig:ar-rf}b,\,f). Examination of the 171\,\AA~emission at this time (panel h) reveals that the footpoint of this loop is located at the outer edge of NRH2. However, at 06:36\,UT, this location resides inside NRH2, i.e., inside the flux rope. This means that the corresponding coronal loop has reconnected. RD images reveal that after 06:21\,UT, the loop contracts towards the erupting filament (panels c and g), and is no longer discernible at 06:28\,UT (panel d).

The footpoint location of this loop (green circle) with respect to the ribbons is also shown in Figure \ref{Fig:aia1600}a. At 06:21\,UT (blue), it is located on the opposite side of NHR2 compared to the violet circle. Thus, loops from both sides of the growing NRH2 reconnect with the erupting flux rope. This is consistent with the recent extensions to the 3D model as analyzed in \citet{Aulanier19} from one MHD simulation of \citet{Zuccarello15}: Initially, the footpoints of the coronal arcades surround the hook (blue in Figure \ref{Fig:ar-rf_model}a), while the erupting flux rope (pink) is rooted inside the hook. The fixed footpoints of the field lines plotted are in Figure \ref{Fig:ar-rf_model} shown by circles. As the eruption progresses, the hook grows, while the straight part of the ribbon moves away from the PIL. The footpoints of the previous flux rope field lines become footpoints of the flare loops (pink in Figure \ref{Fig:ar-rf_model}b), while the footpoints of arcades turn to footpoints of the flux rope (blue). Note however that in the model, there is only one ribbon hook, while in our event, the NRH1 and NRH2 are likely parts of a single deformed hook (see Section \ref{Sect:2}). Finally, the evolution of ribbons due to the ar--rf reconnection geometry is also captured by the cartoon presented in Figure 2 of \citet{Zemanova19}.

Note also that in our event, the filament itself did not seem to reach the location of the green circle as the field lines in Figure \ref{Fig:ar-rf_model} do. The filament itself likely does not represent the entirety of the erupting flux rope, since filaments are only located in the lower, dipped portions of long, flat, sheared, or twisted field lines \citep[e.g.,][]{Gibson06,Dudik08,Luna12,Zuccarello16,Xia16,Gibson18,Gunar18}, while the rest of the flux rope may not be readily apparent in observations. However, as the green circle is located in a coronal dimming region (compare Figure \ref{Fig:aia1600}a with \ref{Fig:aia131}d), the corresponding coronal loop likely reconnected with a flux rope field line not made visible in observations by absorbing filament plasma.

%
%
\section{Late RR--RF Reconnection \\Under the Erupting Filament}
\label{Sect:4}

We also study the occurrence of the rr--rf reconnection in our event. To do that, we use the AIA 94\,\AA~observations, which show morphology similar to 131\,\AA, but are less saturated in the impulsive and peak phases of the flare. 

In the impulsive phase, bright filament strands are observed at the inner side of its legs. At 06:29\,UT, a pair of bright strands start to converge towards one another. The strand in the northern filament leg, denoted by red dotted line, is visible sooner than the conjugate strand within the southern leg, denoted by orange dotted line (Figure \ref{Fig:rr-rf}a--c). The convergence of these bright strands happens directly above the flare arcade and lasts until about 06:31:20\,UT (panel d). Although an X-type geometry is suggestive, this is not an inflow of coronal material (aa--rf reconnection) as reported previously \citep[e.g.,][]{Yokoyama01,Zhu16}. At 06:31:20\,UT, a bright new flare loop appears above the flare arcade, retracting slightly with time (orange arrow in Figure \ref{Fig:rr-rf}f--h). Finally, a bright blob, a remnant of the upper part of the reconnecting southern bright strand, is seen ascending into the filament (white arrows), while the rest of the strand disappears completely (panel f).

Based on the timing and the observed morphology, both of which are consistent with the predictions of the 3D extensions to the standard model, we conclude that we observe the rr--rf reconnection: In accordance with the model, the rr--rf reconnection happens only when the eruption is well developed, i.e., the flux rope legs are stretched vertically on both sides of the central part of the flaring PIL.

Since the new flare loops resulting from the rr--rf reconnection are formed above the existing flare arcade, we speculate that the rr--rf reconnection could be a candidate source for the supra-arcade downflowing loops \citep[e.g.,][]{Savage12a,Savage12b}. Note that we do not observe any supra-arcades in this event to unfavorable on-disk geometry.

%
%
\section{Summary}
\label{Sect:5}

We analyze \textit{SDO}/AIA imaging observations of the filament eruption of 2011 June 07 and report on signatures of the theoretically predicted 3D ar--rf and rr--rf reconnection geometries \citep{Aulanier19} involving the erupting flux rope. These reconnection geometries are found in the well-known eruption of a solar filament observed on 2011 June 07.

According to the theoretical predictions, in the \textit{ar--rf} reconnection geometry, the erupting magnetic flux rope \textit{r} reconnects with a coronal arcade \textit{a} to produce a new flux rope field line \textit{r} and a flare loop \textit{f}.  In our event, the ar--rf reconnection manifests itself in several ways. First, the filament leg shifts its position by more than 25$\arcsec$ northward, to a position previously occupied by a footpoint of a coronal arcade (loop), while the original position of the filament leg becomes a flare loop. The ar--rf reconnection is also associated with the evolution of the flare ribbons and their hooks: as the straight part of the observed ribbon moves away from the polarity inversion line, it narrows down the hook from the inside. Meanwhile the hook is growing, and footpoints of several coronal arcades from both sides of the hook NRH2 become inside the hook. All this is consistent with the evolution of the ribbon hook in the 3D extensions to the standard solar flare model.

The \textit{rr--rf} reconnection geometry involves two legs of the erupting flux rope \textit{r} and produces a new flare loop \textit{f} and a new flux rope field line \textit{r}. In our event, the observed signatures of the rr--rf reconnection consist of the convergence of two bright filament strands in an apparently X-type geometry. This convergence is followed by disappearance of the converging strands and appearance of a new flare loop near the leg of the southern strand, while a bright blob, a remnant of one of the strands, is observed to be ascending into the filament.

Thus, for the first time, both types of reconnections are identified in a single event and in a complete manner, with all four constituents of both the ar--rf and rr--rf reconnections being visible. These observations and their locations are consistent with the recent predictions of 3D extensions to the standard solar flare model \citep{Aulanier19}. Further support for these theoretical results comes from the observed timing with relation to the evolution of the flare. The ar--rf reconnection occurs before the fast eruption and the impulsive phase of the flare, while the rr--rf occurs during the peak phase under the erupting filament. These observations show that the early to peak phases of an eruption are much richer and varied in the reconnection geometries than previously realized, with consequences for evolution of CMEs. 

These observations of the 3D reconnection geometries, together with the evidences for ar--rf reconnection presented by \citet{Zemanova19} and \citet{Lorincik19}, represent the latest successful tests of the predictions of the 3D extensions to the standard flare model. Together with the previous ones \citep{Dudik14a,Dudik16,Dudik17a,Zuccarello17,Aulanier19}, these observational tests contribute to the theoretical advancements in the study of the nature of solar flares and eruptions. They also provide new tools to link interplanetary CME flux ropes with their footpoints at the Sun's surface, in the context of the upcoming Solar Orbiter mission.

\acknowledgments
The authors thank the anonymous referee for comments that helped to improve the manuscript. JD, JL, and AZ acknowledge support from Grant No. 17-16447S of the Grant Agency of the Czech Republic, as well as institutional support RVO:67985815 from the Czech Academy of Sciences. The simulation used in this work was executed on the HPC center MesoPSL which is financed by the R\'egion Ile-de-France and the project Equip@Meso of the PIA supervised by the ANR. G.A. thanks the CNES and the Programme National Soleil Terre of the CNRS/INSU for financial support, as well as the Astronomical Institute of the Czech Academy of Sciences in Ond\v{r}ejov for financial support and warm welcome during his visits.
The authors also thank F. Kotou\v{c}ek and R. Tomek for providing the most excellent coffee one can get in Prague. AIA and HMI data are courtesy of NASA/SDO and the AIA and HMI science teams. 
\facilities{SDO.}

\bibliographystyle{aasjournal}
\bibliography{2019_3D_Reconnection}

\end{document}